%% file: paper.tex
\documentclass[runningheads]{llncs}
\usepackage{graphicx}
\usepackage{appendix}
\usepackage{booktabs}   
\usepackage{subcaption} 

\usepackage{cite}
\usepackage{setspace}            
\usepackage{color}
\usepackage[normalem]{ulem}
\usepackage{url}
\usepackage{amsmath}
\usepackage{multicol}

\usepackage[hypertexnames=false]{hyperref}

\DeclareUrlCommand\ULurl{
  
  }

\usepackage{algorithm}
\PassOptionsToPackage{noend}{algpseudocode}
\usepackage{algpseudocode}

\usepackage{etoolbox}
\usepackage{tikz}
\usetikzlibrary{shapes,snakes}
\usetikzlibrary{patterns}
\usetikzlibrary{tikzmark}
\usetikzlibrary{calc}

\errorcontextlines\maxdimen

\newcommand{\ALGtikzmarkcolor}{black}
\newcommand{\ALGtikzmarkextraindent}{4pt}
\newcommand{\ALGtikzmarkverticaloffsetstart}{-.5ex}
\newcommand{\ALGtikzmarkverticaloffsetend}{-.5ex}
\makeatletter
\newcounter{ALG@tikzmark@tempcnta}

\newcommand\ALG@tikzmark@start{%
    \global\let\ALG@tikzmark@last\ALG@tikzmark@starttext%
    \expandafter\edef\csname ALG@tikzmark@\theALG@nested\endcsname{\theALG@tikzmark@tempcnta}%
    \tikzmark{ALG@tikzmark@start@\csname ALG@tikzmark@\theALG@nested\endcsname}%
    \addtocounter{ALG@tikzmark@tempcnta}{1}%
}

\def\ALG@tikzmark@starttext{start}
\newcommand\ALG@tikzmark@end{%
    \ifx\ALG@tikzmark@last\ALG@tikzmark@starttext
    \else
        \tikzmark{ALG@tikzmark@end@\csname ALG@tikzmark@\theALG@nested\endcsname}%
        \tikz[overlay,remember picture] \draw[\ALGtikzmarkcolor] let \p{S}=($(pic cs:ALG@tikzmark@start@\csname ALG@tikzmark@\theALG@nested\endcsname)+(\ALGtikzmarkextraindent,\ALGtikzmarkverticaloffsetstart)$), \p{E}=($(pic cs:ALG@tikzmark@end@\csname ALG@tikzmark@\theALG@nested\endcsname)+(\ALGtikzmarkextraindent,\ALGtikzmarkverticaloffsetend)$) in (\x{S},\y{S})--(\x{S},\y{E});%
    \fi
    \gdef\ALG@tikzmark@last{end}%
}

\apptocmd{\ALG@beginblock}{\ALG@tikzmark@start}{}{\errmessage{failed to patch}}
\pretocmd{\ALG@endblock}{\ALG@tikzmark@end}{}{\errmessage{failed to patch}}
\makeatother

\algnewcommand{\IfThenElse}[3]{
  \State \algorithmicif\ #1\ \algorithmicthen\ #2\ \algorithmicelse\ #3}
  
\algdef{SE}[DOWHILE]{Do}{doWhile}{\algorithmicdo}[1]{\algorithmicwhile\ #1}%

\algrenewcommand\algorithmicforall{\textbf{foreach}}
\algrenewcommand\algorithmicindent{.8em}

\algnewcommand{\algorithmicforeach}{\textbf{for each}}
\algdef{SE}[FOR]{ForEach}{EndForEach}[1]
  {\algorithmicforeach\ #1\ \algorithmicdo}
  {\algorithmicend\ \algorithmicforeach}

\makeatletter  
\let\OldStatex\Statex
\renewcommand{\Statex}[1][3]{%
  \setlength\@tempdima{\algorithmicindent}%
  \OldStatex\hskip\dimexpr#1\@tempdima\relax}
\makeatother

\newcommand{\casfunc}{\mbox{\sc Cas}}
\newcommand{\ack}{\mbox{\texttt{ACK}}}
\newcommand{\emptyret}{\mbox{\texttt{EMPTY}}}

\newcommand{\pwb}{\mbox{\texttt{pwb}}}
\newcommand{\pfence}{\mbox{\texttt{pfence}}}
\newcommand{\psync}{\mbox{\texttt{psync}}}

\newcommand{\func}[1]{\mbox{\sc #1}}
\newcommand{\com}{\mbox{$\triangleright$}}

\newcommand{\remove}[1]{}

\begin{document}
\title{Flat-Combining-Based Persistent Data Structures for Non-Volatile Memory}

%
%
\author{Matan Rusanovsky\inst{1,3} \and
Hagit Attiya\inst{2} \and
Ohad Ben-Baruch\inst{1} \and Tom Gerby\inst{1} \and Danny Hendler\inst{1} \and Pedro Ramalhete \inst{4}}
\authorrunning{Rusanovsky et al.}
%
\institute{Ben-Gurion University
\email{\{matanru, ohadben, tomger\}@post.bgu.ac.il} 
\email{hendlerd@cs.bgu.ac.il}\and
Department of Computer Science, Technion, Israel \email{hagit@cs.technion.ac.il}
\and Israel Atomic Energy Commission \and Cisco Systems \email{pramalhe@gmail.com}}
\maketitle              
\begin{abstract}
\emph{Flat combining} (FC) is a synchronization paradigm in which a single thread, holding a global lock, collects requests by multiple threads for accessing a concurrent data structure and applies their combined requests to it. Although FC is sequential, it significantly reduces synchronization overheads and cache invalidations and thus often provides better performance than that of lock-free implementations. 

The recent emergence of non-volatile memory (NVM) technologies increases the interest in the development of \textit{persistent} 
concurrent objects. 
These are objects that are able to recover from system failures and ensure consistency by retaining their state in NVM and fixing it, if required, upon recovery. Of particular interest are \emph{detectable} objects that, in addition to ensuring consistency, allow recovery code to infer if a failed operation took effect before the crash and, if it did, obtain its response. 

In this work, we present the first FC-based persistent object implementations. Specifically, we introduce a detectable FC-based implementation of a concurrent LIFO stack, a concurrent FIFO queue, and a double-ended queue. Our empirical evaluation establishes that due to flat combining, the novel implementations require a much smaller number of costly persistence instructions than competing algorithms and are therefore able to significantly outperform them.

\end{abstract}


%
%
%

\input{Introduction}
\input{model}
\input{RFC-Stack}

\input{Memory}
\section{Outline of correctness proof for DFC-Stack} \label{proof-dfc-stack}

\textbf{Linearizability.} Consider an execution of DFC. Denote the value of $cEpoch$ in the beginning of the latest combining phase as $e$. Assume process $p$ was the first process to win $cLock$ as the combiner in $cEpoch=e$. We next provide the linearization points for all threads that submitted an operation, in the following two cases: 1) $p$'s first $cEpoch$ modification in line \ref{first-inc} was not persisted to the NVM, 2) $p$'s first $cEpoch$ modification in line \ref{first-inc} was successfully persisted to the NVM, i.e. $cEpoch$ stores $e'>e$ in the persisted memory. 

As long as the first case holds, no operation of any thread will be linearized. 
However when the second case holds, all threads that were collected in the current epoch number, i.e. threads for which the if statement in line \ref{if-op-ready-for-collect} was evaluated as true for the combiner, are linearized immediately when the first change to $cEpoch$ is persisted to the NVM. 

Note that the second case is possible if we assume that from some point in time, threads do not fail-stop, and therefore eventually there will be an epoch in which threads will be allowed to complete their execution before the system fails. Then, after some additional crash-free steps of the system, the epoch number will be incremented again and all collected threads will be able to retrieve their response values.

First all the collected combined operations are linearized, followed by all other collected operations, in an increasing order of their IDs - which is the order in which the combiner collects them in line \ref{loop-reduce-first}. If a couple of push and pop operations were combined in \func{Reduce}, then the pop operation will be linearized immediately after the push operation. Any other threads that finished submitting their jobs after the combiner concluded collecting the submitted operations of the current epoch, will not be linearized. These threads are considered as late arriving threads, and they are guaranteed to be collected and linearized in the next crash-free combining phase.

In case of a crash while the first case holds ($p$'s first $cEpoch$ modification was not persisted to the NVM), in \mbox{\sc Recover} another process $q$ wins $cLock$ and becomes the recovery combiner. The same logic as above applies to this case - until case 2 is fulfilled, while substituting the recovery combiner thread $q$ with $p$.
Note that crashes while case 2 holds do not break linearizability, as operations from the last combining phase that were linearized just after the change to $cEpoch$ was persisted will not be collected again because their response value field will not be reset in line \ref{reset-val-recover}, and therefore will not be empty in line \ref{if-op-ready-for-collect}.

\textbf{Starvation-freedom for crash-free executions.} We assume for the progress condition, that starting from some point there are no more system failures or, at least, the system does not crash again until all threads complete their current operations. Otherwise, no progress can be established if the system crashes before threads complete their operations.

Consider the crash-free execution that begins from the point above. Note that as long as a thread submits an operation, after a finite number of steps (of the combiner, or the next combiner if this thread is late arriving), its operation is guaranteed to be applied, from similar arguments to FC \cite{hendler2010flat}. It is left to notice that after a finite number of steps, (line \ref{set-next-op-collect}) each thread finishes to submit its new operation.

\input{Evaluation}

\input{Discussion}

\section*{Acknowledgments}
This work was supported by Pazy grant 226/20.
Computational support was provided by the NegevHPC project~\cite{negevhpc}.

%
%
%
\bibliographystyle{splncs04}
\bibliography{mybibliography}


\appendix

\section{DFC-based Queue and Deque}
\label{queue and deque}

\subsection{Code for DFC Queue and Deque}

Based on the DFC stack shown before we implemented a DFC based queue and deque. 
These objects follows closely the DFC algorithm.
both are represented by a linked list.

\paragraph{DFC Queue} implement a FIFO queue that allows two operations - enqueue and dequeue.
To perform them a thread first announce them. Then, it tries to capture the lock and become a \emph{combiner}. If it succeeds, it proceeds to execute $\func{Combine}$; otherwise, it waits in $\func{LockTaken}$ until its operation is applied by the combiner. Opposed to the Stack the DFC Queue don't use a reduce procedure to cancel pairs of operations, The cases which allow the FIFO queue to perform cancels are limited reducing the benefits we saw in the case of the stack.
Upon recovery from a crash, 
each resurrected thread executes the $\mbox{\sc Recover}$ procedure which remains without changes from the stack algorithm. The queue algorithm and data structures are further presented in figure \ref{alg:types2} and algorithm \ref{dfc-queue}.

\paragraph{DFC Deque} an object supporting both FIFO and LIFO operations - pushFront, popFront allows to push and pop nodes from the front side of the list.  pushRear, popRear allows to push and pop nodes from the rear side of the list.
To perform them a thread first announce them. Then, it tries to capture the lock and become a \emph{combiner}. If it succeeds, it proceeds to execute $\func{Combine}$; otherwise, it waits in $\func{LockTaken}$ until its operation is applied by the combiner.
The combiner first traverses an announcement array and eliminates 
pairs of \func{Push} and \func{Pop} on the same side of the queue - for example pairs of \func{PushFront} and \func{PopFront} are canceled via the $\func{Reduce}$ procedure. 
Then, it commits the surplus announced push or pop operations 

Upon recovery from a crash, 
each resurrected thread executes the $\mbox{\sc Recover}$ procedure which remains without changes from the stack algorithm.The deque algorithm and data structures are further presented in figure \ref{alg:types3} and algorithms \ref{deque-1} \ref{deque-2}.

\subsection{DFC Queue and Deque experimental results}

\paragraph{Experimental Setting and Benchmarks}
Our experiments were conducted using a machine \remove{\cite{negevhpc}} equipped with two sockets of Intel Xeon Gold 5215 processors, with a cache size of 13.75 MB each. It contains a total of 20 physical and 40 logical cores and 4 256GB Intel Optane Persistent Memory Modules configured to the App Direct mode. The machine contains also 192 GB of RAM and runs on CentOS Linux with kernel version 5.8.7. All algorithms were implemented in C++ and compiled with g++ (version 9.1.0) with O2 optimizations.
On Intel architectures \cite{Performance-Measurements}, the \pwb\ instruction is translated to \texttt{clwb}, \texttt{clflush} or \texttt{clflushopt} depending on what the machine at hand supports, where \texttt{clflushopt} is the most efficient option. In addition, the \pfence\ and \psync\ instructions are translated to \texttt{sfence}. Consequently, our code uses the \texttt{clflushopt} instruction and assumes that the execution of \texttt{sfence} ensures also the functionality of \psync.

In the case of queue - The first benchmark we used is the \textit{enq-deq} benchmark, in which we executed 1M couples of enq and deq operations, that were distributed equally between the threads. In the second benchmark, called \textit{rand-op}, each thread chooses randomly and independently each of its operations to be either a enq or a deq operation. In total, 2M operations are performed, distributed equally between the threads.
We ran the benchmarks using a varying number of threads in order to test algorithms' throughput and scalability. Each experiment was repeated 10 times and we report on median results.

In the case of deque - The first benchmark we used is the \textit{pushf-pushr-popf-popr} benchmark, in which we executed 1M couples of push and pop operations, that were distributed equally between the threads.The side in which the pair is executed front/rear is chosen randomly. In the second benchmark, called \textit{rand-op}, each thread chooses randomly and independently each of its operations to be either a pushFront, popFront, pushRear or popRear operation. In total, 2M operations are performed, distributed equally between the threads.
We ran the benchmarks using a varying number of threads in order to test algorithms' throughput and scalability. Each experiment was repeated 10 times and we report on median results.

\paragraph{Queue Experimental Results}

\begin{figure*}[tb]
	\begin{subfigure}[c]{0.5\textwidth}
		\includegraphics[height=3.9cm]{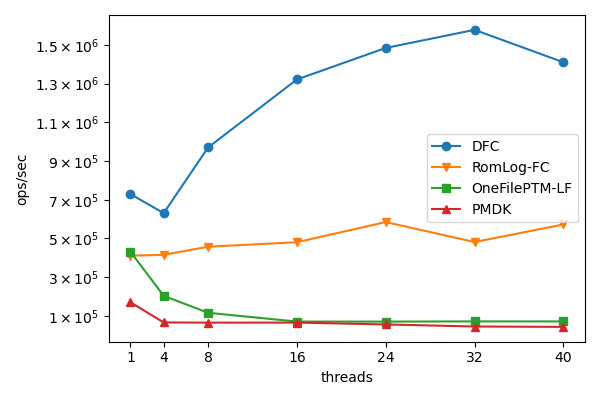}
		\subcaption{enq-deq throughput}
		\label{fig:enq-deq-throughput}
	\end{subfigure}		
	\begin{subfigure}[c]{0.5\textwidth}
		\includegraphics[height=4.1cm]{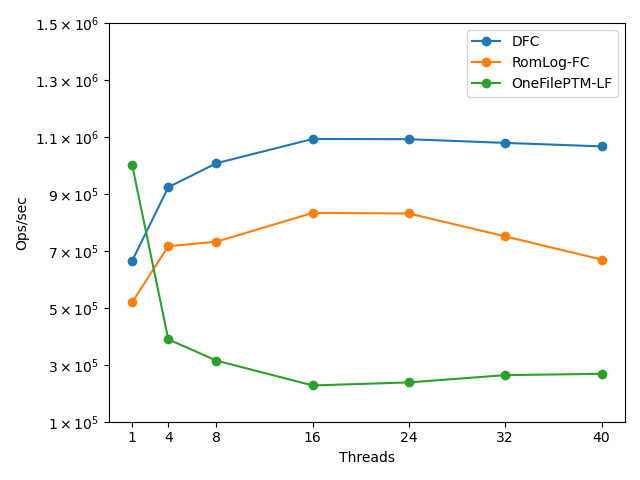}
		\subcaption{rand-op throughput}
		\label{fig:rand-op-queue-throughput}
	\end{subfigure}

	\caption{Queue Evaluation: Throughput for the enq-deq benchmark and rand-op benchmark.}
	\label{fig:graphs2}
\end{figure*}

Graphs are presented in Figure \ref{fig:graphs2}.
As shown by Figure \ref{fig:enq-deq-throughput} and \ref{fig:rand-op-queue-throughput}, \DFC\ and \Romulus\ outperform \OneFile\ and \PMDK\ by a wide margin for all concurrency levels. And \DFC\ outperforms \Romulus\ through all concurrency levels in the queue test. 

\paragraph{Deque Experimental Results}

\begin{figure*}[tb]
	\begin{subfigure}[c]{0.5\textwidth}
		\includegraphics[height=3.9cm]{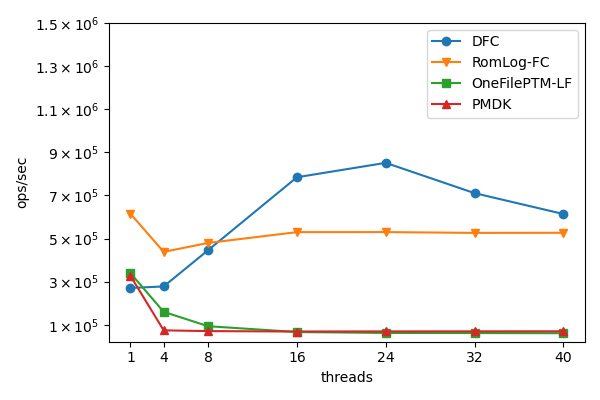}
		\subcaption{pushf/r-popf/r throughput}
		\label{fig:deque-throughput}
	\end{subfigure}		
	\begin{subfigure}[c]{0.5\textwidth}
		\includegraphics[height=4.1cm]{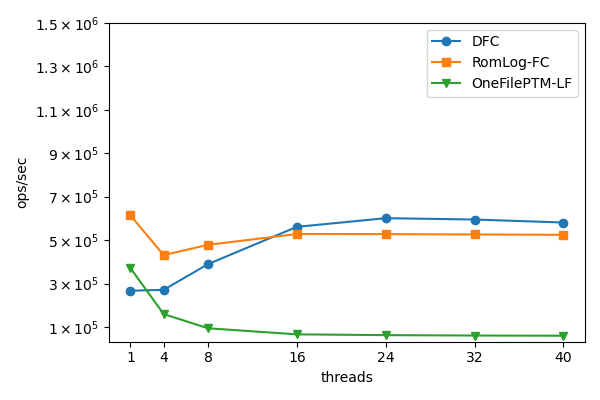}
		\subcaption{rand-op throughput}
		\label{fig:deque-randop-throughput}
	\end{subfigure}		
	
	\caption{Deque Evaluation: Throughput the push-pop benchmark and rand-op benchmark.}
	\label{fig:graphs3}
\end{figure*}

Graphs are presented in Figure \ref{fig:graphs3}.
As shown by Figure \ref{fig:deque-throughput} and \ref{fig:deque-randop-throughput}, \DFC\ and \Romulus\ outperform \OneFile\ and \PMDK\ by a wide margin for all concurrency levels.

\subsection{Outline of correctness proof for DFC Queue and Deque}
The same correctness proof arguments that were presented for the stack object in ~Appendix \ref{proof-dfc-stack} can be applied for the queue and the deque objects, except for few small modifications, specifically in the 2nd case, in which the first $cEpoch$ modification in lines \ref{queue-first-cepoch} and \ref{deque-first-cepoch} respectively was successfully persisted to the NVM. While in the stack all the collected operations (first the combined couples and then all others) were linearized in an increasing order of their IDs, in the queue first all the collected enqueue operations are linearized in an increased order of their IDs, followed by all the collected and ordered dequeue operations. In the deque, first all the collected combined operations on the head of the queue are linearized in an increasing order, and then the collected combined operations on the tail are linearized in an increasing order. Finally all other collected operations on the head, followed by the collected operations on the tail are linearized in an increasing order.

\begin{figure}
	
	\scriptsize		
	\begin{flushleft}
	    
	    \begin{multicols*}{2}
	    type Node \{ \\
	    \hspace*{6mm} Integer $param$ \\
	    \hspace*{6mm} Node* $next$ \\
	    \hspace*{4mm} \} \\
		
		Type TAnnounce \{ \\
		\hspace*{6mm} 2-bit variable $valid$ \\
		\hspace*{6mm} Announce $ann$[2] \\
		\hspace*{4mm} \}
		
		\columnbreak
		
		type Announce \{  \\
		\hspace*{6mm} Integer $val$ \\
		\hspace*{6mm} Integer $epoch$ \\
		\hspace*{6mm} Integer $param$ \\
		\hspace*{6mm} \{\func{Enq}, \func{Deq}\} $name$  \\
		\hspace*{4mm} \} \\
		
		\end{multicols*}
		
		\com\ \textbf{Non-Volatile shared variables:} \\
		\hspace*{4mm} Integer $cEpoch$, initially 0 \\
		\hspace*{4mm} Node* $top$[2], initially both $\perp$ \\
		\hspace*{4mm} Node* $bot$[2], initially both $\perp$ \\
		\hspace*{4mm} TAnnounce $tAnn$[N], initially all fields are 0 \\
		
		\com\ \textbf{Volatile shared variables:} \\
		\hspace*{4mm} CAS variable $cLock$, initially 0 \\
		\hspace*{4mm} CAS variable $rLock$, initially 0 \\
		\hspace*{4mm} Integer $enqList$[N], $deqList$[N], $vColl$[N], initially all 0 \\

	\end{flushleft}	
	\caption{DFC Queue: types and initialization}
	\label{alg:types2}
\end{figure}

\newpage

\begin{algorithm*}
\scriptsize
\caption{\textbf{DFC Queue.} 
Pseudocode for \func{Combine} and \func{Collect} procedures.
}

\label{dfc-queue}


\begin{multicols*}{2}
\begin{algorithmic}[1]
\Procedure{Combine}{ }
    \State $tEnq, tDeq :=$ \mbox{\sc Collect}() 
	\State $ head := top[(cEpoch/2) \% 2]$ 
	\State $ tail := bot[(cEpoch/2) \% 2]$
        \If{$tEnq > 0$ }
            \While {$tEnq > 0$ }
                \State $tEnq--$
                \State $cId := enqList[tEnq]$ 
                \State $vOp := vColl[cId]$
                \State $param := tAnn[cId].ann[vOp].param$ 
		        \State $ nNode :=$ \func{Allocate Node} 
		        \Statex[6] ($param$, $tail$) 
                \State $ tAnn[cId].ann[vOp].val :=$ \ack 
		        \State \pwb($\& nNode$) 
		        \State $tail := nNode$ 
          \EndWhile
          \EndIf
        \If {$tDeq > 0$  }
	        \While {$tDeq > 0$ }
	            \State $tDeq--$
                \State $cId := deqList[tDeq]$
                \State $vOp := vColl[cId]$
		        \If{$head = \perp$}
		            \State $ tAnn[cId].ann[vOp].val :=$ \emptyret 
		        \Else
		            \State $ tAnn[cId].ann[vOp].val :=$ 
		            \Statex[6] $head.param $ 
		            \State $tempHead := head$
		            \State $ head := head.next$
		            \State \mbox{\sc Deallocate Node}($tempHead$)
		            
		            \If{$head = \perp$}
		                \State{$tail = \perp$}
		            \EndIf
		        \EndIf
          \EndWhile
        \EndIf
    \State $top[(cEpoch/2 + 1)\%2] := head$ 
    \State $bot[(cEpoch/2 + 1)\%2] := tail$
	\For {$i=1$ to $N$}
	    \State $vOp := vColl[i]$ 
	    \If {$vOp \neq \perp$} \pwb($\& tAnn[i].ann[vOp]$)  \EndIf
	\EndFor
	\State \pwb($top[(cEpoch/2 + 1)\%2]$);
	\State \pwb($bot[(cEpoch/2 + 1)\%2]$);
	\pfence() 
	\State $cEpoch ++$ \label{queue-first-cepoch}
	\State \pwb($\& cEpoch$); \pfence() 
	\State $cEpoch ++$ 
	\State $cLock := 0$ 
	\State \textbf{return} 
\EndProcedure
\columnbreak

\Procedure{Collect}{ }
	\State $tEnq, tDeq := -1$ 
	\For {$i=1$ to $N$} 
	    \State $vOp := tAnn[t].valid$ 
	    \State $opVal := tAnn[i].ann[LSB(vOp)].val$ 
	    \If{$MSB(vOp) = 1$ and $opVal = \perp$} 
		        \State $tAnn[i].ann[LSB(vOp)].epoch :=$ 
		        \Statex[6] $cEpoch$ 
		        \State $vColl[i] := LSB(vOp)$ 
		        \If{$tAnn[i].ann[LSB(vOp)].name =$ 
		        \Statex[6] $Enqueue$}  
		            \State $tEnq ++$ 
		            \State $enqList[tEnq] := i$ 
		        \Else
		            \State $tDeq ++$ \;
		            \State $deqList[tDeq] := i$ 
		    \EndIf
	    \Else
	        \State $vColl[i] := \perp$ 
        \EndIf
    \EndFor
    
    \State\textbf{return} $tEnq, tDeq$
 \EndProcedure

\end{algorithmic}
\end{multicols*}
\end{algorithm*}

\begin{figure}
	
	\scriptsize		
	\begin{flushleft}
	    
	    \begin{multicols*}{2}
	    type Node \{ \\
	    \hspace*{6mm} Integer $param$ \\
	    \hspace*{6mm} Node* $next$ \\
	    \hspace*{6mm} Node* $prev$ \\
	    \hspace*{4mm} \} \\
		
		Type TAnnounce \{ \\
		\hspace*{6mm} 2-bit variable $valid$ \\
		\hspace*{6mm} Announce $ann$[2] \\
		\hspace*{4mm} \}
		
		\columnbreak
		
		type Announce \{  \\
		\hspace*{6mm} Integer $val$ \\
		\hspace*{6mm} Integer $epoch$ \\
		\hspace*{6mm} Integer $param$ \\
		\hspace*{6mm} \{\func{PushFront}, \func{PushRear}\, \func{PopFront}, \func{PopRear}\} $name$  \\
		\hspace*{4mm} \} \\
		
		\end{multicols*}
		
		\com\ \textbf{Non-Volatile shared variables:} \\
		\hspace*{4mm} Integer $cEpoch$, initially 0 \\
		\hspace*{4mm} Node* $top$[2], initially both $\perp$ \\
		\hspace*{4mm} Node* $bot$[2], initially both $\perp$ \\
		\hspace*{4mm} TAnnounce $tAnn$[N], initially all fields are 0 \\
		
		\com\ \textbf{Volatile shared variables:} \\
		\hspace*{4mm} CAS variable $cLock$, initially 0 \\
		\hspace*{4mm} CAS variable $rLock$, initially 0 \\
		\hspace*{4mm} Integer $pushFrontList$[N], $popFrontList$[N], $vColl$[N], initially all 0 \\
		\hspace*{4mm} Integer $pushRearList$[N], $popRearList$[N], initially all 0 \\

	\end{flushleft}	
	\caption{DFC Deque: types and initialization}
	\label{alg:types3}
\end{figure}

\begin{algorithm*}
\scriptsize
\caption{\textbf{DFC Deque.} 
Pseudocode for \func{Combine} procedure.
}


\label{deque-1}
\begin{multicols*}{2}
\begin{algorithmic}[1]

\Procedure{Combine}{ }
    \State $tIndexFront, tIndexRear :=$ \mbox{\sc Reduce}() 
	\State $ head := top[(cEpoch/2) \% 2]$
	\State $ tail := bot[(cEpoch/2) \% 2]$ 
        \If{$tIndexFront > 0$}
            \While {$tIndexFront > 0$ }
                \State $tIndexFront--$
                \State $cId := pushFrontList[tIndexFront]$ 
                \State $vOp := vColl[cId]$
                \State $param := tAnn[cId].ann[vOp].param$ 
		        \State $nNode := $ \func{Allocate Node}
		        \Statex[6] ($param$, $head$) 
                \State $ tAnn[cId].ann[vOp].val :=$ \ack 
		        \State \pwb($\& nNode$)
		        \If{$head = \perp$}
		            \State $head := nNode$
		            \State $tail := nNode$
		        \Else
		        \State $head := nNode$
		        \EndIf 
          \EndWhile
        \ElsIf {$tIndexFront < 0$ }
            \State $tIndexFront := -1 \cdot tIndexFront$  
	        \While {$tIndexFront > 0$ }
	            \State $tIndexFront--$
                \State $cId := popFrontList[tIndexFront]$
                \State $vOp := vColl[cId]$
		        \If{$head = \perp$}
		            \State $ tAnn[cId].ann[vOp].val :=$ \emptyret 
		        \Else
		            \State $tAnn[cId].ann[vOp].val:=$
		            \Statex[6] $head.param$ 
		            \State $tempHead:= head$
		            \State $head := head.next$
		            \If{$head != \perp$}
		            \State $head.prev := \perp$
		        \Else
		        \State $tail := \perp$
		        \EndIf
		            
		            \State \mbox{\sc Deallocate Node}($tempHead$) 
		        \EndIf
          \EndWhile
        \EndIf
        
        \columnbreak
        
        \If{$tIndexRear > 0$ }
            \While {$tIndexRear > 0$}
                \State $tIndexRear--$
                \State $cId := pushRearList[tIndexRear]$ 
                \State $vOp := vColl[cId]$
                \State $param := tAnn[cId].ann[vOp].param$ 
		        \State $nNode := $ \func{Allocate Node} ($param$) 
                \State $ tAnn[cId].ann[vOp].val :=$ \ack 
		        \State \pwb($\& nNode$) 
		        
		        \If{$tail = \perp$}
		            \State $head := nNode$
		            \State $tail := nNode$
		        \Else
		        \State $nNode.prev := tail$
		        \State $tail.next := nNode$
		        \State $tail := nNode$
		        \EndIf

          \EndWhile
        \ElsIf {$tIndexRear < 0$ }
            \State $tIndexRear := -1 \cdot tIndexRear$  
	        \While {$tIndexRear > 0$}
	            \State $tIndexRear--$
                \State $cId := popRearList[tIndexRear]$
                \State $vOp := vColl[cId]$
		        \If{$tail = \perp$}
		            \State $ tAnn[cId].ann[vOp].val :=$ \emptyret
		        \Else
		            \State $tAnn[cId].ann[vOp].val:=$
		            \Statex[6] $tail.param$ 
		            \If{$tail.prev = \perp$}
		            \State $head := \perp$
		            \State $tail := \perp$
		        \Else
		        \State $tail := tail.prev$
		        \State $tail.next := \perp$
		        \EndIf

		            \State \mbox{\sc Deallocate Node}($tempHead$) 
		        \EndIf
          \EndWhile
        \EndIf
        
    \State $top[(cEpoch/2 + 1)\%2] := head$ 
    \State $bot[(cEpoch/2 + 1)\%2] := head$ 
	\For {$i=1$ to $N$}
	    \State $vOp := vColl[i]$ 
	    \If {$vOp \neq \perp$} \pwb($\& tAnn[i].ann[vOp]$)  \EndIf
	\EndFor
	\State \pwb($top[(cEpoch/2 + 1)\%2]$)
	\State \pwb($bot[(cEpoch/2 + 1)\%2]$); \pfence() 
	\State $cEpoch ++$  \label{deque-first-cepoch}
	\State \pwb($\& cEpoch$); \pfence() 
	\State $cEpoch ++$ 
	\State $cLock := 0$ 
	\State \textbf{return} 
\EndProcedure

\end{algorithmic}
\end{multicols*}
\end{algorithm*}

\begin{algorithm*}
\scriptsize
\caption{\textbf{DFC Deque.} 
Pseudocode for \func{Reduce} procedure.
}


\label{deque-2}

\begin{multicols*}{2}
\begin{algorithmic}[1]

\Procedure{Reduce}{ }
	\State $tPushFront, tPopFront := -1$ 
	\State $tPushRear, tPopRear := -1$
	\State $tIndexRear, tIndexFront := -1$ 
	\For {$i=1$ to $N$} 
	    \State $vOp := tAnn[t].valid$
	    \State $opVal := tAnn[i].ann[LSB(vOp)].val$ 
	    \If{$MSB(vOp) = 1$ and $opVal = \perp$ } 
	        \State $tAnn[i].ann[LSB(vOp)].epoch :=$ 
	        \Statex[6] $cEpoch$ 
	        \State $vColl[i] := LSB(vOp)$ 
	        \If{$tAnn[i].ann[LSB(vOp)].name =$ 
	        \Statex[6]$PushFront$}  
	            \State $tPushFront ++$ 
	            \State $pushFrontList[tPushFront] := i$ 
	        \Else
	             \If{$tAnn[i].ann[LSB(vOp)].name =$
	             \Statex[6]$PushRear$} 
    	            \State $tPushRear ++$ \;
    	            \State $pushRearList[tPushRear] := i$ 
	            \Else
	         \If{$tAnn[i].ann[LSB(vOp)].name =$
	         \Statex[6]$PopRear$} 
	            \State $tPopRear ++$ \;
	            \State $popRearList[tPopRear] := i$
	        \Else
	            \State $tPopFront ++$ \;
	            \State $popFrontList[tPopFront] := i$
	        \EndIf
	        \EndIf
	    \EndIf
	    \Else
	        \State $vColl[i] := \perp$ 
        \EndIf
    \EndFor
    \item[]
    \item[]
	\While {$tPushFront \neq -1$ and 
	\Statex[6] $tPopFront \neq -1$ }
		    \State $cPush := pushFrontList[tPushFront]$ 
			\State $cPop := popFrontList[tPopFront]$ 
			\State $vPush := vColl[cPush]$ 
	        \State $tAnn[cPush].ann[vPush].val :=$ \ack 
	        \State $vPop := vColl[cPop]$ 
	        \State $tAnn[cPop].ann[vPop].val := $
	        \Statex[6] $ tAnn[cPush].ann[vPush].param$ 
	        \State $tPushFront--$
	        \State $tPopFront--$
    \EndWhile
    \While {$tPushRear \neq -1$ and 
    \Statex[6] $tPopRear \neq -1$ }
		    \State $cPush := pushRearList[tPushRear]$ 
			\State $cPop := popRearList[tPopRear]$ 
			\State $vPush := vColl[cPush]$ 
	        \State $tAnn[cPush].ann[vPush].val :=$ \ack 
	        \State $vPop := vColl[cPop]$ 
	        \State $tAnn[cPop].ann[vPop].val := $
	        \Statex[6] $ tAnn[cPush].ann[vPush].param$ 
	        \State $tPushRear--$
	        \State $tPopRear--$
    \EndWhile
    
    \If {$tPushFront \neq -1 $}  
    \State $tIndexFront = tPushFront + 1$
	\EndIf
	\If {$tPopFront \neq -1 $}  
	\State $tIndexFront = -1 * (tPopFront + 1)$
	\EndIf
	\If {$tPushRear \neq -1 $}  
	\State $tIndexRear = tPushRear + 1$
	\EndIf
	\If {$tPopRear \neq -1 $}
	\State $tIndexRear = -1 * (tPopFront + 1)$
	\EndIf
	
	\State \textbf{return} $tIndexRear, tIndexFront$

 \EndProcedure

\end{algorithmic}
\end{multicols*}
\end{algorithm*}

\end{document}

%% file: Introduction.tex
\section{Introduction}
\label{sec-introduction}

Byte-addressable non-volatile main memory (NVM) combines the performance
beneﬁts of conventional (volatile) DRAM-based main memory with the durabil-
ity of secondary storage. The recent availability of NVM-based systems increased
the interest in the development of \emph{persistent}
concurrent objects. These are objects that are able to recover from system failures (\emph{crashes}) and ensure consistency by retaining their state in NVM and fixing it, if required, upon recovery. Of particular interest are \emph{detectable} objects \cite{FriedmanHMP18} that, in addition to ensuring consistency, allow recovery code to infer if a failed operation took effect before the crash and, if it did, obtain its response. 

The correctness condition for persistent objects that we use in this work is \emph{durable linearizability}~\cite{IzraelevitzMS16}, which, simply stated, requires that linearizability \cite{HerlihyW90} be maintained in spite of crash-failures. 
Devising durably linearizable recoverable objects in general, and detectable ones in particular, is challenging. Although data stored in main memory will not be lost upon a system crash, with the currently available technology, caches and registers \emph{are} volatile and their content \emph{is} lost if the system fails before it is \emph{persisted} (that is, written to NVM). Since operations on concurrent objects are, in general, not atomic, a system crash may occur in the midst of operations applied to the object and leave it in an inconsistent state that must be fixed upon recovery. Ensuring correctness is made even more complicated due to the fact that cache lines are not necessarily evicted in the order in which they were written by the program. Consequently, program stores may be persisted out of order. Persistence order can be guaranteed by explicitly invoking \emph{persistence instructions} such as flushes and fences. However, these instructions are expensive and should be used as sparingly as possible.



A key approach for devising persistent concurrent objects is  by using memory transactions. \emph{Persistent transactional memories} (PTMs) (e.g. \cite{correia2018romulus,kolli2016high,marathe2018persistent,ramalhete2019onefile,pmdksite,volos2011mnemosyne,wu2019architecture}) are general-purpose implementations that support persistent memory transactions. A PTM implementation supports the execution of concurrent transactions by multiple threads while ensuring that the effect of each transaction is either persisted as a whole or has no effect. Although PTMs make the construction of persistent objects easier for programmers, they often incur significant performance overheads, largely because they have to maintain metadata for ensuring transactional semantics. Moreover, as we show in this work, they also incur extra persistence instructions in comparison with an optimized implementation of a specific concurrent object that is able to leverage object semantics. Another shortcoming of PTMs is that, to the best of our knowledge, none of them provides detectability. Unfortunately, for many key concurrent objects, optimized non-transactional detectable implementations still do not exist.


\emph{Flat combining} (\emph{FC})~\cite{hendler2010flat} is 
a coarse-grained lock-based synchronization technique,
in which threads delegate their work to a single \emph{combiner thread},
which combines operations by multiple threads in a manner that 
exploits the semantics of the implemented concurrent object and then jointly applies them.
For example, several \emph{add} operations on a counter can be replaced with
a single \emph{add} operation with the sum of their arguments. 
Informally, FC proceeds in \emph{combining phases}, and in each phase,
threads contend on a single global lock; 
the winner becomes the combiner for that phase, 
while the rest of the threads wait for their operation to be completed by the combiner. 
Despite being sequential, the reduced synchronization overhead and cache invalidations 
of FC more than offset the higher parallelism of alternative implementations. 
\cite{hendler2010flat} presented (non-persistent) FC-based implementations of stacks, queues
and priority queues that outperform prior implementations of these objects that 
are either lock-based~\cite{yew1987distributing,shalev2006predictive}  
or nonblocking~\cite{scott2001scalable,hoffman2007baskets,
shavit2000skiplist,HendlerSY10}.   

This work presents \emph{detectable flat combining} (\emph{DFC}),
an approach for persistent objects that is based on FC,
and applies it to derive detectable stacks, queues and double-ended queues. 
We experimented in our DFC objects with the concept of elimination. In the case of a stack object, pairs of  \emph{push} and \emph{pop} are combined by ``eliminating'' them: 
each \emph{pop} operation in the pair return the item that is the argument 
of the \emph{push} operation in the pair. 
Surplus \emph{push} or \emph{pop} operations are combined by atomically 
extending or truncating a linked-list stack representation. 
As shown by our experimental evaluation, when the combiner thread collects a large number 
of pairs that are suitable for elimination, performance is significantly improved. 

\paragraph{Contributions.}

\begin{itemize}
\item We present DFC stack, a flat-combining implementation of a 
persistent and detectable stack; 
it is the first non-transactional detectable persistent stack. 
To the best of our knowledge, 
DFC is the also the first flat-combining-based implementation of \emph{any} persistent object.

\item We experimentally evaluate the DFC stack on an Intel machine with NVM.
We compare its performance with that of data structures implemented on top 
of the Romulus~\cite{correia2018romulus}, OneFile~\cite{ramalhete2019onefile} 
and Intel's PMDK PTM~\cite{pmdksite} . 
The DFC stack outperforms all these implementations by a wide margin. 
Specifically, DFC leverages the stack semantics to significantly reduce 
the number of persistence instructions in comparison with the other algorithms.

\item 
The key algorithmic ideas we use for making the FC technique persistent 
are generic. We apply DFC to derive detectable persistent queues and double-ended queues (deques). 
Experimental evaluation shows that DFC-based queues and deques provide
good performance. The pseudo-code and evaluation of our DFC-based queue and deque are presented in the appendix.

\end{itemize}



%% file: model.tex
\section{System Model}
\label{sec-model}


We assume the \emph{shared cache model}~\cite{IzraelevitzMS16},
in which the shared memory holds both 
non-volatile shared variables (residing in NVM) 
and volatile shared variables (residing in DRAM). 
The contents of the cache and processor registers are volatile:
primitive operations are applied to volatile memory, 
since they are applied to variables residing in registers or in the cache.
Writes to non-volatile variables are persisted to NVM using explicit
flush instructions, or when a cache line is evicted.
Under \emph{explicit epoch persistency}~\cite{IzraelevitzMS16},
a write-back to persistent storage is triggered by a
\emph{persistent write-back (\pwb)} instruction.
The thread may continue its execution after \pwb{} is issued.
The order of \pwb{} instructions is not necessarily preserved.
When ordering is required, a \pfence{} instruction orders 
preceding \pwb{} instructions before all subsequent \pwb{} instructions.
A \psync{} instruction waits until all previous \pwb{} instructions
complete the write back. 
For each memory location, persistent write-backs preserve program order. 

We assume the \emph{Total Store Order} (\emph{TSO}) model,
supported by the x86 and SPARC architectures,
where writes become visible in program order.
In addition, since in current architectures supporting NVM a \pfence{} 
acts as both \pfence{} and \psync, 
our pseudocode uses \pfence{} to indicate the execution of both.

At any point during the execution of an operation, 
a \emph{system-wide crash-failure} (or simply a \textit{crash}) may occur, 
which resets all volatile variables to their initial values, 
but preserves the values stored in the NVM.
An operation's response is lost if a crash occurs before it was 
\emph{persisted} to a non-volatile variable.

Following a crash, the system resurrects all threads and lets them execute the $\texttt{Recover}$ procedure, in order to recover the data-structure by fixing inconsistencies in it, if any. The system may crash again before resurrecting all threads. Thus, $\texttt{Recover}$ may be invoked multiple times before it completes, because the system may undergo multiple crashes in the course of executing it.

DFC ensures \emph{durable linearizability}~\cite{IzraelevitzMS16},
that is, linearizability is maintained despite crashes:
once the system recovers after a crash-failure,
the state of the data structure reflects a history containing all operations 
that completed before the crash and 
may also contain some operations that have not completed before the crash. 
This captures the idea that an operation must be linearized 
only once its effect is persisted to NVM.
An implementation is \emph{detectable}~\cite{ben2019delay} if $\texttt{Recover}$ 
also finishes $p$'s crashed operation (if there is one) and returns its response.

DFC provides \emph{starvation-freedom}: 
If a thread invokes an operation \textit{Op}, 
all active threads continue taking steps and the system does not crash, 
then eventually \textit{Op} completes,
either directly or by the completion of $\texttt{Recover}$.

%% file: RFC-Stack.tex
\section{The DFC Stack} \label{dfc_desc}

In this section, we describe the DFC stack algorithm. 
As done in FC-based algorithms, each process \emph{announces} its operation (in the DFC stack algorithm, this is either a \func{Push} or a \func{Pop}), by writing its operation code and arguments to its entry  in an \emph{announcement array}. Then, each process attempts to capture a global lock that protects a sequential data structure and become a \emph{combiner}. Processes that fail to capture the lock (\emph{non-combiner} processes) wait for the combiner to apply their operations, whereas the combiner proceeds to traverse the announcement array, collect announced operations, apply them to the data structure, and write operation responses to their corresponding announcement array entries. In our implementation, the stack is represented by a linked list.

In a plain-vanilla FC-based stack implementation, the combiner applies \func{Push} and \func{Pop} operations by simply adding or removing a list item from/to the head of the list. In our implementation, the combiner also employs (whenever possible) \emph{elimination} \cite{HendlerSY10}, to pair concurrent \func{Push} and \func{Pop} operations. The combiner applies each such operations pair by setting the response of the \func{Pop} operation to be the input of the \func{Push} operation without having to access the linked list. The combiner needs to modify the linked list only if the numbers of \func{Push} and \func{Pop} operations that it collected differ. If there is a surplus of non paired-up \func{Push} operations, the combiner appends nodes containing their arguments to the head of the list; if there is a surplus of \func{Pop} operations, then the respective number of nodes is removed from the list (if they exist) and their values are written to the response fields of the respective announcement array; if the surplus of \func{Pop} operations is larger than the length of the list, some \func{Pop} operations return an empty response value. We now describe the algorithm in more detail.

Data types, shared variables and initialization values are presented in Figure \ref{alg:types}.
Variable $cEpoch$ is a global epoch counter that counts the number of combining phases that have been performed so far (multiplied by 2, for reasons we explain soon). $top$ is a two-entry array that stores two pointers, such that after each combining phase, in an alternating manner, one of them points to the head of the stack. The current epoch number, stored in $cEpoch$, is used in order to determine which of the $top$ pointers is the up-to-date head of the linked list representing the stack. $tAnn$ is the announcement array. It contains $N$ entries, each of which stores an $ann$ array consisting of two announcement structures. Each announcement structure has fields for storing the operation announcement, response, and a 2-bit variable $valid$, whose least-significant-bit (LSB) indicates which of the announcement structures is the active one, and whose most-significant-bit (MSB) is set only once the announcement is ready for the combiner to collect. For simplicity of presentation, we use LSB($valid$) and MSB($valid$) to access and set the respective bit of $valid$. $cLock$ is the combiner lock. The rest of the variables presented in Figure \ref{alg:types} are described when they are first mentioned in the pseudocode description that follows.

\begin{figure}[tb]
	
	\scriptsize		
	    
	    \begin{multicols*}{2}
	    type Node \{ \\
	    \hspace*{6mm} Integer $param$ \\
	    \hspace*{6mm} Node* $next$  \} 
		
		type TAnnounce \{ \\
		\hspace*{6mm} 2-bit variable $valid$ \\
		\hspace*{6mm} Announce $ann$[2]  \}

		type Announce \{  \\
		\hspace*{6mm} Integer $val$ \\
		\hspace*{6mm} Integer $epoch$ \\
		\hspace*{6mm} Integer $param$ \\
		\hspace*{6mm} \{\func{Push}, \func{Pop}\} $name$  \} \\

		\columnbreak
        \com\ \textbf{Non-Volatile shared variables:} \\
		\hspace*{4mm} Integer $cEpoch$, initially 0 \\
		\hspace*{4mm} Node* $top$[2], initially both $\perp$ \\
		\hspace*{4mm} TAnnounce $tAnn$[N], all initially 0 \\
		
		\com\ \textbf{Volatile shared variables:} \\
		\hspace*{4mm} CAS variable $cLock$, initially 0 \\
		\hspace*{4mm} CAS variable $rLock$, initially 0 \\
		\hspace*{4mm} Integer $pushList$[N], $popList$[N], \\
		\hspace*{14mm} $vColl$[N], all initially  0 \\

		\end{multicols*}
	\caption{DFC Stack: types and initialization}
	\label{alg:types}
\end{figure}


Algorithm \ref{dfc}(left) presents the pseudocode of the \func{Op} procedure, which implements both \func{Push} and \func{Pop} operations.
At the beginning of each operation (lines \ref{read-cepoch-op}-\ref{fix-opepoch-op}), a thread $t$ first creates a local copy $opEpoch$ of the current $cEpoch$ and checks if it is even. If it is not, $t$ increments $opEpoch$ to the next even number. 
Then, in lines \ref{get-next-op}-\ref{write-op-name}, $t$ announces its operation in the next available announcement structure (the one not currently pointed at by $valid$). It does so by writing the operation type and argument (the latter only in case of a \func{Push}), resetting the response value in $val$ to a special value $\perp$ and updating the structure's $epoch$ field.
The following \pwb\ and \pfence\ instructions ensure that the values in the announcement structure are persisted before the $valid$ field is modified to point to the updated announcement structure (line \ref{set-next-op}). This ensures that in case of a crash, $valid$ does not point to the wrong announcement structure. Then, \pwb\ and \pfence\ instructions are executed again to ensure that $valid$ is persisted, so that in case of a crash the \emph{recovery combiner} will handle the correct announcement structure. After $valid$ is persisted, $t$ sets the MSB of $valid$ in line \ref{set-next-op-collect}, notifying the combiner that its announcement is ready to be combined.

\begin{algorithm*}
\scriptsize
\caption{\textbf{DFC Stack.} 
\func{Op}, \func{Recover} and auxiliary procedures}
\label{dfc}
\begin{multicols*}{2}

\begin{algorithmic}[1]
\makeatletter
\makeatother
\Procedure{Op}{$param$}
	\State $opEpoch := cEpoch$ \label{read-cepoch-op}
	\If {$opEpoch \% 2 = 1$} {$opEpoch++$} \label{fix-opepoch-op} \EndIf
	\State $nOp := 1 - LSB(tAnn[t].valid) $ \label{get-next-op}
	\State $tAnn[t].ann[nOp].val := \perp $ \label{write-res-val}
	\State $tAnn[t].ann[nOp].epoch := opEpoch $ \label{write-epoch-val}
	\State $tAnn[t].ann[nOp].param := param $ \label{write-op-param}
	\State $tAnn[t].ann[nOp].name := Op $ \label{write-op-name}
	\State \pwb($\& tAnn[t].ann[nOp]$); \pfence() \label{parallel1}
	\State $tAnn[t].valid = nOp $ \label{set-next-op} 
	\State \pwb($\& tAnn[t].valid$); \pfence() \label{parallel2}
	\State $MSB(tAnn[t].valid) := 1 $ \label{set-next-op-collect} 
	\State $value := $ \func{TakeLock}($opEpoch$) \label{InvokeTakeLock}
	\If {$value \neq \perp$} 
	    \State \textbf{return} $value$
	\Else
	    \State \func{Combine}() \label{combine-op}
	    \State \textbf{return} $tAnn[t].ann[nOp].val$ \label{combiner-return}
    \EndIf
\EndProcedure

\Procedure{TakeLock}{$opEpoch$}
\label{try-to-take-lock}
	\If{$cLock.\casfunc\ (0, 1) = False$ } 
		\While {$cEpoch \leq opEpoch+1$\label{spin}}
		    \If{($cLock=0$ and
		    \Statex[6] $cEpoch\leq opEpoch+1$) \label{try-to-take-lock-from-spin}}
		        \State \textbf{return} \func{TakeLock} ($opEpoch$) 
		    \EndIf
		\EndWhile
		\State \textbf{return} \func{TryToReturn}($opEpoch$)
	\Else {} \textbf{return} $\perp$
	\EndIf
\EndProcedure

\remove{
\Procedure{TryToTakeLock}{$opEpoch$}
\label{try-to-take-lock}
	\State $combiner := cLock.\casfunc\ (0, 1)$ 
	\State \textbf{return} \mbox{\sc LockTaken}($opEpoch, combiner$)
\EndProcedure

\Procedure{LockTaken}{$opEpoch, combiner$}
\label{lock-taken}
	\If{$combiner = False$ \label{not-combiner}} 
		\While {$cEpoch \leq opEpoch+1$\label{spin}}
		    \If{$cLock=0$ and $cEpoch \leq opEpoch+1$ \label{try-to-take-lock-from-spin}}
		        \State \textbf{return} \mbox{\sc TryToTakeLock}($opEpoch$) 
		    \EndIf
		\EndWhile
		\State \textbf{return} \mbox{\sc TryToReturn}($opEpoch$)
		\Else
		    \State \textbf{return} $\perp$
	\EndIf
\EndProcedure
}
			
\columnbreak
			
\Procedure{Recover}{$param$}	
	\If{$rLock.\casfunc\ (0, 1)$ \label{take-recovery-lock}}
		\If {$cEpoch \% 2 = 1$ \label{odd_epoch}}
            \State $cEpoch ++$ \label{increment}
            \State \pwb($\& cEpoch$); \pfence() \label{fix_cepoch_pwb_pfence}
        \EndIf
	    \State \mbox{\sc GarbageCollect}() \label{gc-recovery}
		\For {$i=1$ to $N$} 
		    \State $vOp := tAnn[i].valid$ 
		    \State $opEpoch := tAnn[i].ann[LSB(vOp)].epoch$
		    \If{$MSB(vOp) = 0$}
		        \State $MSB(tAnn[i].valid) := 1$ \label{set-valid-recover}
		    \EndIf
		    \If{$opEpoch = cEpoch$ \label{if-opepoch-is-cepoch}}
		        \State $tAnn[i].ann[LSB(vOp)].val := \perp$ \label{reset-val-recover}
	        \EndIf
		\EndFor
		\State \mbox{\sc Combine}() \label{combine-recover}
		\State $rLock = 2$ \label{release-recovery-lock}
	\Else
	    \While {$rLock = 1$ \label{recovery-spin}} spin \EndWhile
	\EndIf
	\State \textbf{return} $tAnn[t].ann[LSB(tAnn[t].valid)].val$ \label{return-val-recovery}
\EndProcedure

\Procedure{TryToReturn}{$opEpoch$}
	\State $vOp := LSB(tAnn[t].valid) $ \label{get-res-valid-op}
	\State $val := tAnn[t].ann[vOp].val$ \label{get-res-val}
	\If{$val = \perp$ \label{get-value-after-cepoch-inc}} \algorithmiccomment{late arrival}
	    \State $opEpoch := opEpoch + 2$ \label{epoch-add-2}
	    \State \textbf{return} \func{TakeLock}($opEpoch$)
	\Else { } \textbf{return} $val$ \label{check-if-returned}
	\EndIf
\EndProcedure

\algstore{part1}
\end{algorithmic}
\end{multicols*}
\end{algorithm*}

\begin{algorithm*}
\scriptsize
\caption{\textbf{DFC Stack.} 
\func{Combine} and \func{Reduce} procedures.
}
\label{combine}

\begin{multicols*}{2}
\begin{algorithmic}[1]
\algrestore{part1}
\Procedure{Combine}{ }
    \State $tIndex :=$ \mbox{\sc Reduce}() \label{reduce-ops}
	\State $ head := top[(cEpoch/2) \% 2]$ 
        \If{$tIndex > 0$ \label{surplus-push-ops}}
            \While {$tIndex > 0$ \label{there-is-push}}
                \State $tIndex--$
                \State $cId := pushList[tIndex]$ \label{get-collected-push}
                \State $vOp := vColl[cId]$
                \State $param := tAnn[cId].ann[vOp].param$ 
		        \State $nNode :=$ \func{Allocate Node}
		        \Statex[6] ($param$, $head$) \label{allocate-new-node}
                \State $ tAnn[cId].ann[vOp].val :=$ \ack \label{ret-val-push}
		        \State \pwb($\& nNode$) \label{pwb-nNode}
		        \State $head := nNode$ \label{push-end}
          \EndWhile
        \ElsIf {$tIndex < 0$  \label{surplus-pop}}
            \State $tIndex := -1 \cdot tIndex$  \label{fix-index-pop}
	        \While {$tIndex > 0$ \label{there-is-pop}}
	            \State $tIndex--$
                \State $cId := popList[tIndex]$
                \State $vOp := vColl[cId]$
		        \If{$head = \perp$}
		            \State $ tAnn[cId].ann[vOp].val :=$ \emptyret \label{pop-empty}
		        \Else
		            \State $tAnn[cId].ann[vOp].val:=$
		            \Statex[6] $head.param$ \label{pop-something}
		            \State $tempHead:= head$
		            \State $head := head.next$
		            \State \mbox{\sc Deallocate Node}($tempHead$) \label{deallocate}
		        \EndIf
          \EndWhile
        \EndIf
    \State $top[(cEpoch/2 + 1)\%2] := head$ 
	\For {$i=1$ to $N$}
	    \State $vOp := vColl[i]$ 
	    \If {$vOp \neq \perp$} \pwb($\& tAnn[i].ann[vOp]$)\label{PWB-announce} \EndIf
	\EndFor
	\State \pwb($top[(cEpoch/2 + 1)\%2]$); \pfence() 
	\State $cEpoch ++$ \label{first-inc}
	\State \pwb($\& cEpoch$); \pfence() \label{first-inc-persist}
	\State $cEpoch ++$ \label{second-inc}
	\State $cLock := 0$ \label{free-lock} 
	\State \textbf{return} 
\EndProcedure

\columnbreak

\Procedure{Reduce}{ }
	\State $tPush, tPop := -1$ 
	\For {$i=1$ to $N$ \label{loop-reduce-first}} 
	    \State $vOp := tAnn[t].valid$ \label{get-valid-op}
	    \State $opVal := tAnn[i].ann[LSB(vOp)].val$ 
	    \If{$MSB(vOp) = 1$ and $opVal = \perp$ \label{if-op-ready-for-collect}} 
		        \State $tAnn[i].ann[LSB(vOp)].epoch :=$
		        \Statex[6]$cEpoch$ \label{update-old-epoch}
		        \State $vColl[i] := LSB(vOp)$ \label{vc-update}
		        \If{$tAnn[i].ann[LSB(vOp)].name$
		        \Statex[6] $= Push$}  
		            \State $tPush ++$ 
		            \State $pushList[tPush] := i$ \label{push-pushlist}
		        \Else
		            \State $tPop ++$ \;
		            \State $popList[tPop] := i$ \label{push-poplist}
		    \EndIf
	    \Else
	        \State $vColl[i] := \perp$ \label{vc-no}
        \EndIf
    \EndFor
	\While {$tPush \neq -1$ and $tPop \neq -1$ \label{actual_reduce}}
		    \State $cPush := pushList[tPush]$ 
			\State $cPop := popList[tPop]$ 
			\State $vPush := vColl[cPush]$ \label{get-push-valid-op}
	        \State $tAnn[cPush].ann[vPush].val :=$ \ack \label{imm-push}
	        \State $vPop := vColl[cPop]$ \label{get-pop-valid-op}
	        \State $tAnn[cPop].ann[vPop].val := $
	        \Statex[6] $ tAnn[cPush].ann[vPush].param$\label{imm-pop}
	        \State $tPush--$
	        \State $tPop--$
    \EndWhile
	\If {$tPush \neq -1 $} \textbf{return} $tPush + 1$
	\EndIf
	\If {$tPop \neq -1 $} 
	\Statex[6]\textbf{return} $-1 * (tPop + 1)$
	\EndIf
	\State \textbf{return} $0$ \label{epmty-list}
 \EndProcedure
 
\end{algorithmic}
\end{multicols*}
\end{algorithm*}

The update of $valid$ is done in this manner in order to deal with the following bad scenario. Suppose a non-combiner thread completes announcing its new operation and changes $valid$ to point to it, but is yet to persist it. Now, a combiner collects the operation and applies it. In case of a crash, $valid$ may point at the old announcement structure (since it was not persisted). Thus, the thread will not be able to tell whether its current operation was completed or not. An alternative solution to this issue is to have the combiner itself persist the $valid$ field of any operation it collects. However, this would be degrade the performance, since the combiner delays all other threads, and we opted not to do so. Note also that  persistence instructions performed by the announcement code may be executed by different threads in parallel, so a \pfence\ instruction by one thread does not block the progress of other threads. In contrast, persistence instructions performed by the combiner delay all waiting non-combiner threads. As shown by our evaluation results, the adverse effect of these concurrent persistence instructions on the performance of DFC is smaller than that of the combiner's persistence instructions. 


After announcing its operation, a thread attempts to become the combiner by capturing the combiner lock $cLock$ in the \func{TakeLock} procedure (line \ref{InvokeTakeLock}). We first describe the combiner code, and discuss the code performed by non-combiners later. The combiner returns from \func{TakeLock} without waiting and proceeds to combine all announced operations in line \ref{combine-op}, by calling the \mbox{\sc Combine} procedure (Algorithm \ref{combine}). 
Only a single thread (the current combiner) may execute this procedure at any given time. 
\func{Combine} first calls \mbox{\sc Reduce} in line \ref{reduce-ops}, which is responsible to combine all announced operations and pair-up \func{Push} and \func{Pop} operations. Each such pair of  operations is eliminated by setting the response of the \func{Pop} operation to the argument of the corresponding \func{Push} operation; the response of the \func{Push} operation simply indicates that it was completed. 

More specifically, \func{Reduce} iterates over the $tAnn$ array and collects all announcements that are ready, as indicated by the $valid$ field, and that contain no response value in their $val$ field (line \ref{if-op-ready-for-collect}). 
For each collected operation, the combiner updates (in line \ref{update-old-epoch}) the epoch in the announcement structure to be the current global epoch number, thus indicating that the operation was collected (and is about to be applied) during the current combining phase. This is required for dealing with the following potential problem. In case of a crash during a combining phase, as we explain in the description of the \mbox{\sc Recover} procedure below, the algorithm collects again all the operations of the failed combining phase. This is done because the algorithm may have paired up a \func{Push} and a \func{Pop} operations together before the crash but may have updated the response only for one of them. For example, the response of a \func{Pop} operation might have been  set to the value field of the paired \func{Push} operation, but the response of that \func{Push} operation might not have been updated before the crash, hence it might be coupled with a second \func{Pop} operation after the crash, thus violating correctness. 
A similar problem may arise for operations that are applied by accessing the linked-list representing the stack. Consequently, in such cases, the recovery combiner must re-apply all operations and therefore needs to re-collect all the announcements of the crashed combining phase, regardless of their response values. 
In order to allow identifying operations that must be re-collected, each announcement structure stores the epoch in which it was collected by the combiner.

\remove{
The combiner then updates in line \ref{update-old-epoch} the epoch of the worker to the current epoch number, so to indicate for late-arrivals that they are being combing in the current epoch, and to deal with the following issue. In case of a crash during the current combine, as we explain in the description of the \mbox{\sc Recover} procedure below, we would like to collect and re-execute all operations again. A problem might happen if we paired a push and a pop operations together before the crash, and were able to update only one of them about their completion. This might violate the consistency of the data-structure. For example the paired pop operation returns that its return value is the value of the pushed parameter of the paired push operation. However, the paired push operation might only later be coupled with, for instance, another pop operation.
For that, each operation should hold the epoch in which the combiner has collected it.}

The combiner inserts operations into pop and push operation lists -- to be latter combined and applied. Since there are at most $N$ \func{Push} and \func{Pop} operations, these lists are implemented using volatile arrays of size $N$.
We also use a volatile array $vColl$ of size $N$, indicating for each thread whether its operation has been collected and, if so, storing the index of the active announcement structure (line \ref{vc-update}). If no operation was collected for a thread, the corresponding entry is set to $\perp$ (line \ref{vc-no}).
In the loop of line \ref{actual_reduce}, the combiner performs the actual reduction/elimination by eliminating couples of \func{Push} and \func{Pop} operations, as long as none of the lists is exhausted, and updates their responses directly without accessing the linked list stack representation. This elimination reduces the number of \pwb\ instructions, since accessing the stack requires additional \pwb s, to persist the changes made to it.
Finally, \func{Reduce} returns the surplus push or pop operations. It does so by returning the number of surplus operations left. A positive number indicates surplus \func{Push} operations, while a negative number indicates surplus \func{Pop} operations. This way, the combiner can later access these operations using the appropriate operations list.

\func{Combine} then applies the surplus operations, returned by  \func{reduce}, to the stack. The while loop in line \ref{there-is-push} deals with the case of surplus \func{Push} operations, and the while loop in line \ref{there-is-pop} deals with the case of surplus \func{Pop} operations. 
For each \func{Push} operation, a new non-volatile node is allocated and added to the central stack, representing the actual operation. In addition, \ack\ is updated as the response value.
Likewise, for each \func{Pop} operation, a node is removed from the stack (if one exists), its key is used as the response value (line \ref{pop-something}), and finally the node is de-allocated.
After applying all operations to the stack, the next $top$ entry -- $(cEpoch/2+1)\%2$, is set to point to the new head of the stack. If there was no surplus, the new $top$ entry equals the previous $top$ entry.
Finally, the combiner persists all modified variables, specifically, all the combined announcement structures (line \ref{PWB-announce}) and the updated $top$ entry, using \pwb\ instructions for all these variables, followed by \emph{a single} \pfence\ instruction. After ensuring that the stack and all responses are in a consistent and persistent state, $cEpoch$ is incremented twice and persisted once in between these two updates. Then, the combiner releases the combiner lock in line \ref{free-lock} and \func{Combine} returns. The operation of the combiner is guaranteed to have been completed, thus the thread returns its response in line \ref{combiner-return}.

$cEpoch$ is incremented twice in order to deal with the following scenario. 
Consider a configuration in which the combiner increments $cEpoch$ from $v$ to $v+1$, at the end of a combining phase, and then a crash occurs before the combiner manages to persist $v+1$.
Suppose additionally that some thread $t$ managed to observe the updated $cEpoch$ before the crash. Thus, $t$ observed a return value and can conclude that its operation was completed, therefore $t$ may ``\textit{safely}'' complete its operation and begin a new one. 
However, after the crash, the recovery function may observe the old $cEpoch$ value $v$ again, concluding that $t$'th operation was not performed. Moreover, this is indistinguishable from the case in which a crash occurred in the middle of the combining phase, \emph{before} $cEpoch$ was updated. However, in the first case it is not safe to re-execute $t$'th operation again during recovery, since $t$ may have started a new one, while in the second case $t$'s operation must be executed upon recovery, as it did not take effect.
To solve this issue, we release non-combiner threads only when they observe that $cEpoch$ was incremented twice (lines \ref{spin}--\ref{try-to-take-lock-from-spin}), thus ensuring that the announced operations were linearized properly and that the combiner successfully persisted $v+1$ in line \ref{first-inc-persist}. In case of a crash, if the value of $cEpoch$ is reverted back to an odd number $v+1$, it is safe to consider the epoch phase $v$ as completed, and simply fix $cEpoch$ to $v+2$. For this reason, we only persist $cEpoch$ after updating it to $v+1$, but not after updating it to $v+2$.

We next describe what non-combiner threads do while waiting for their response. In line \ref{spin}, a non-combiner busy-waits for the combiner to increase $cEpoch$ by 2 (or more). If this condition is satisfied, it is guaranteed that all combined announcement structures received valid response values. Consequently, if the non-combiner thread finds a response value in line \ref{get-value-after-cepoch-inc}, it can safely return that value. Otherwise, if $cEpoch$ was incremented but there is no response value, the thread has arrived late, that is, it has completed announcing its operation only after the combiner checked its announcement structure. 
In this case, $opEpoch$ is incremented by two in line \ref{epoch-add-2} in order to wait for the next combiner to collect the operation. The algorithm guarantees that this scenario can occur only once, since the next combiner will surely collect the thread's announcement. 
Note that if a non-combiner announces an operation after the combiner increases the epoch counter but before it releases the lock, that thread might spin forever waiting for a combiner in line \ref{spin}. Line \ref{try-to-take-lock-from-spin} deals with this scenario, by attempting to capture the combiner lock again in case the lock is released and the epoch number was not incremented. 
To facilitate better understanding of the data structure used by DFC, we illustrate below two configurations reached in the course of its execution. In the first configuration, depicted by  Figure \ref{fig:steady}, the current epoch number is 10 but all announcement structures were last updated in previous combining phases. There is no active combiner in this configuration. Each thread is associated with a 2-bit entry in the $valid$ array, whose LSB indexes a two-entry array of announcement structures (shown below the $valid$ array). The LSB of the $valid$ entry indicates which is the active announcement structure of the corresponding thread. The MSB of the $valid$ entry indicates whether the announcement is complete (i.e. valid). A combiner should combine an announcement if it is active, the MSB of the corresponding $valid$ entry is set, and the \emph{val} field of the announcement was not yet set (i.e. it equals $\bot$). Announcement entries that are colored in grey correspond to active announcements. The dotted announcement entry of $p_4$ is still not ready, as indicated by the MSB bit of the corresponding $valid$ entry. Consequently, it was not previously combined and thus its return value was not set yet.
The two-entry array $top$ stores two pointers to the stack below. 
The current entry of $top$ is determined by the value $(cEpoch/2)\%2$, so the second cell, colored in grey, is the current entry of $top$ in this configuration.

Figure~\ref{fig:combine} depicts a second configuration, in which a combining phase is ongoing. The announcement structures of threads $p_1, p_2, p_4$ and $p_5$ were combined. Announcement structures colored in dark grey correspond to valid announcements that were combined in the current combining phase. In this configuration, the combiner, thread $p_4$, already updated the response values of the operations by $p_1,p_2,p_4$, but still did not update the response value of $p_5$'s operation. Here, the combiner eliminated the operations of $p_1$ and $p_2$ without accessing the stack, updated the epoch number and the response value of its own operation and added a corresponding node to the stack. As we described previously, if the system crashes in this configuration (before $cEpoch$ is incremented), all these operations must be combined and applied again.

\input{example_figure}

\paragraph*{The Recovery Procedure.}
In case of a crash, all threads execute the \mbox{\sc Recover} procedure upon recovery (Algorithm~\ref{dfc}(right)). 
If required, the recovery function recovers the shared stack by re-executing the last combining phase, thus completing all pending operations and updating their responses. 
Each thread first attempts to capture the recovery lock $rLock$ that protects the critical section of the recovery code. If the thread fails to capture the lock, it simply busy-waits until the lock is freed in line \ref{recovery-spin}. If it succeeds, it becomes the \textit{recovery combiner}. 
The recovery combiner increments the current epoch number to an even number (line~\ref{odd_epoch}), in case there was a crash after the first $cEpoch$ increment was persisted but before the second increment was persisted (lines \ref{first-inc}-\ref{second-inc}). Note that it is safe to simply increment $cEpoch$ to the next even number, since, as was previously explained, if an odd value  of $cEpoch$ was persisted, all operations from the previous epoch phase, $v$, were already linearized and persisted.
In order to maintain the consistency and durability of $cEpoch$ for all future arriving threads, a \pwb\ and \pfence\ instructions are executed in line \ref{fix_cepoch_pwb_pfence}.
The recovery combiner then fixes inconsistencies in the memory layout by performing garbage collection (line~\ref{gc-recovery}), as explained in Section~\ref{sec-memory} which describes memory management in our algorithm.

Then all announcement structures are traversed again. For each thread, the MSB of the $valid$ field is set in line \ref{set-valid-recover}. This allows the recovery combiner to later combine and apply the operation, if needed. The if statement in lines \ref{if-opepoch-is-cepoch}-\ref{reset-val-recover} is needed for operations that were applied during the last crashed combining phase for which response values were persisted before the crash. In this case, DFC needs to re-collect and re-apply these operations during the recovery. 
Note that the epoch number of operations that received a response value in the last combining phase was updated to the current epoch number in line \ref{update-old-epoch}. Since both the response field $val$ and the epoch number field $epoch$ reside in the same aligned cache line of 64 bytes, our machine's architecture guarantees that they are flushed together to the persistent memory. Consequently, if the response value was persisted before the crash, so was the epoch number. Thus, after the execution of line \ref{reset-val-recover}, all the announcement structures from the last combining phase have empty response values.
We emphasize that even if this guarantee does not hold for other NVM-based machines, one could simply add a \pwb\ instruction after the write to the $epoch$ field in \func{Reduce}.

The recovery combiner proceeds by starting another combining phase in line \ref{combine-recover}. All operations from the last crashed combining phase will be collected and applied again in the \func{Reduce} function. 
We note that threads that did not manage to complete the announcement of their operations before the crash are not re-executed. Finally, the recovery combiner releases the lock in line \ref{release-recovery-lock} and returns its own response value (as do all other threads) in line \ref{return-val-recovery}.

Our DFC queue and deque algorithms are very similar to that of the DFC stack, except that we do not employ elimination in our queue implementation because this technique is more suitable for LIFO than for FIFO order. We provide the pseudo-codes of these implementations in Appendix~\ref{queue and deque}.

%% file: example_figure.tex
\begin{figure*}
	\centering
	\begin{subfigure}[b]{0.7\textwidth}

    	\begin{tikzpicture}
    	\begin{scope}[scale=0.85, transform shape]
    	
    	\draw[fill=black!10] (-0.8+1,-1.5) rectangle (0.5+1,-0.5) node[label={[label distance=0.25cm, minimum height=0cm]}, pos=.5] {};
    	\draw[] (-0.8+1,-0.5) rectangle (0.5+1,0.5) node[label={[label distance=0.25cm, minimum height=0cm]}, pos=.5] {};
    	\draw[fill=black!10] (0.6+1,-1.5) rectangle (1.9+1,-0.5) node[label={[label distance=0.25cm, minimum height=0cm]}, pos=.5] {};
    	\draw[] (0.6+1,-0.5) rectangle (1.9+1,0.5) node[label={[label distance=0.25cm, minimum height=0cm]}, pos=.5] {};
    	\draw[] (2+1,-1.5) rectangle (3.3+1,-0.5) node[label={[label distance=0.25cm, minimum height=0cm]}, pos=.5] {};
    	\draw[fill=black!10] (2+1,-0.5) rectangle (3.3+1,0.5) node[label={[label distance=0.25cm, minimum height=0cm]}, pos=.5] {};
    	\draw[pattern=north west lines, pattern color=black!40] (3.4+1,-1.5) rectangle (4.7+1,-0.5) node[label={[label distance=0.25cm, minimum height=0cm]}, pos=.5] {};
    	\draw[] (3.4+1,-0.5) rectangle (4.7+1,0.5) node[label={[label distance=0.25cm, minimum height=0cm]}, pos=.5] {};
    	\draw[] (4.8+1,-1.5) rectangle (6.1+1,-0.5) node[label={[label distance=0.25cm, minimum height=0cm]}, pos=.5] {};
    	\draw[fill=black!10] (4.8+1,-0.5) rectangle (6.1+1,0.5) node[label={[label distance=0.25cm, minimum height=0cm]}, pos=.5] {};
    	
    	
    	\node[circle, minimum height=1.1cm, dotted] (we) at (-0.15+1,0.3) {\footnotesize{$\func{Pop}()$}};
    	\node[circle, minimum height=1.1cm, dotted] (we) at (1.25+1,0.3) {\footnotesize{$\func{Push}(b)$}};
    	\node[circle, minimum height=1.1cm, dotted] (we) at (2.65+1,0.3) {\footnotesize{$\func{Push}(c)$}};
    	\node[circle, minimum height=1.1cm, dotted] (we) at (4.05+1,0.3) {\footnotesize{$\func{Pop}()$}};
    	\node[circle, minimum height=1.1cm, dotted] (we) at (5.45+1,0.3) {\footnotesize{$\func{Push}(e)$}};
    	
    	\node[circle, minimum height=1.1cm, dotted] (we) at (-0.15+1,0) {\footnotesize{$2$}};
    	\node[circle, minimum height=1.1cm, dotted] (we) at (1.25+1,0) {\footnotesize{$4$}};
    	\node[circle, minimum height=1.1cm, dotted] (we) at (2.65+1,0) {\footnotesize{$4$}};
    	\node[circle, minimum height=1.1cm, dotted] (we) at (4.05+1,0) {\footnotesize{$4$}};
    	\node[circle, minimum height=1.1cm, dotted] (we) at (5.45+1,0) {\footnotesize{$8$}};
    	
    	\node[circle, minimum height=1.1cm, dotted] (we) at (-0.15+1,-0.3) {\footnotesize{\emptyret}};
    	\node[circle, minimum height=1.1cm, dotted] (we) at (1.25+1,-0.3) {\footnotesize{\ack}};
    	\node[circle, minimum height=1.1cm, dotted] (we) at (2.65+1,-0.3) {\footnotesize{\ack}};
    	\node[circle, minimum height=1.1cm, dotted] (we) at (4.05+1,-0.3) {\footnotesize{$b$}};
    	\node[circle, minimum height=1.1cm, dotted] (we) at (5.45+1,-0.3) {\footnotesize{\ack}};
    	
    	\draw[-, anchor=north] (-0.8+1,-0.5) -- node[above left, scale=0.8, sloped] {} (0.5+1,-0.5);
    	\draw[-, anchor=north] (0.6+1,-0.5) -- node[above left, scale=0.8, sloped] {} (1.9+1,-0.5);
    	\draw[-, anchor=north] (2+1,-0.5) -- node[above left, scale=0.8, sloped] {} (3.3+1,-0.5);
    	\draw[-, anchor=north] (3.4+1,-0.5) -- node[above left, scale=0.8, sloped] {} (4.7+1,-0.5);
    	\draw[-, anchor=north] (4.8+1,-0.5) -- node[above left, scale=0.8, sloped] {} (6.1+1,-0.5);
    	
    	\node[circle, minimum height=1.1cm, dotted] (we) at (-0.15+1,-0.7) {\footnotesize{$\func{Push}(d)$}};
    	\node[circle, minimum height=1.1cm, dotted] (we) at (1.25+1,-0.7) {\footnotesize{$\func{Pop}()$}};
    	\node[circle, minimum height=1.1cm, dotted] (we) at (2.65+1,-0.7) {\footnotesize{$\func{Push}(a)$}};
    	\node[circle, minimum height=1.1cm, dotted] (we) at (4.05+1,-0.7) {\footnotesize{$\func{Push}(f)$}};
    	\node[circle, minimum height=1.1cm, dotted] (we) at (5.45+1,-0.7) {\footnotesize{$\func{Pop}()$}};
    	
    	\node[circle, minimum height=1.1cm, dotted] (we) at (-0.15+1,-1) {\footnotesize{$6$}};
    	\node[circle, minimum height=1.1cm, dotted] (we) at (1.25+1,-1) {\footnotesize{$8$}};
    	\node[circle, minimum height=1.1cm, dotted] (we) at (2.65+1,-1) {\footnotesize{$2$}};
    	\node[circle, minimum height=1.1cm, dotted] (we) at (4.05+1,-1) {\footnotesize{$8$}};
    	\node[circle, minimum height=1.1cm, dotted] (we) at (5.45+1,-1) {\footnotesize{$2$}};
    	
    	\node[circle, minimum height=1.1cm, dotted] (we) at (-0.15+1,-1.3) {\footnotesize{\ack}};
    	\node[circle, minimum height=1.1cm, dotted] (we) at (1.25+1,-1.3) {\footnotesize{$d$}};
    	\node[circle, minimum height=1.1cm, dotted] (we) at (2.65+1,-1.3) {\footnotesize{\ack}};
    	\node[circle, minimum height=1.1cm, dotted] (we) at (4.05+1,-1.3) {\footnotesize{$\bot$}};
    	\node[circle, minimum height=1.1cm, dotted] (we) at (5.45+1,-1.3) {\footnotesize{$a$}};
    	
    	\node[circle, minimum height=1.1cm, dotted] (we) at (-0.15+1,-1.75) {$p_1$};
    	\node[circle, minimum height=1.1cm, dotted] (we) at (1.25+1,-1.75) {$p_2$};
    	\node[circle, minimum height=1.1cm, dotted] (we) at (2.65+1,-1.75) {$p_3$};
    	\node[circle, minimum height=1.1cm, dotted] (we) at (4.05+1,-1.75) {$p_4$};
    	\node[circle, minimum height=1.1cm, dotted] (we) at (5.45+1,-1.75) {$p_5$};
    	
    	\node[circle, minimum height=1.1cm, dotted] (we) at (-1.5+1.3,0.3) {\footnotesize{\func{Op}}};
    	\node[circle, minimum height=1.1cm, dotted] (we) at (-1.5+1.3,0) {\footnotesize{epoch}};
    	\node[circle, minimum height=1.1cm, dotted] (we) at (-1.5+1.3,-0.3) {\footnotesize{val}};
    	\node[circle, minimum height=1.1cm, dotted] (we) at (-1.5+1.3,-0.7) {\footnotesize{\func{Op}}};
    	\node[circle, minimum height=1.1cm, dotted] (we) at (-1.5+1.3,-1) {\footnotesize{epoch}};
    	\node[circle, minimum height=1.1cm, dotted] (we) at (-1.5+1.3,-1.3) {\footnotesize{val}};

    	\draw[] (0.2+1,0.8) rectangle (5+1,1.3) node[label={[label distance=2.65cm]}, pos=.5] {};
    	\draw[-, anchor=north] (0.2+0.96*1+1,0.8) -- node[above left, scale=0.8, sloped] {} (0.2+0.96*1+1,1.3);
    	\draw[-, anchor=north] (0.2+0.96*2+1,0.8) -- node[above left, scale=0.8, sloped] {} (0.2+0.96*2+1,1.3);
    	\draw[-, anchor=north] (0.2+0.96*3+1,0.8) -- node[above left, scale=0.8, sloped] {} (0.2+0.96*3+1,1.3);
    	\draw[-, anchor=north] (0.2+0.96*4+1,0.8) -- node[above left, scale=0.8, sloped] {} (0.2+0.96*4+1,1.3);
    	
    	\node[circle, minimum height=1.1cm, dotted] (we) at (-1.5+1.3,1.05) {\small{valid}};

    	\node[circle, minimum height=1.1cm, dotted] (we) at (0.68+1,1.05) {$11$};
    	\node[circle, minimum height=1.1cm, dotted] (we) at (0.68+0.96*1+1,1.05) {$11$};
    	\node[circle, minimum height=1.1cm, dotted] (we) at (0.68+0.96*2+1,1.05) {$10$};
    	\node[circle, minimum height=1.1cm, dotted] (we) at (0.68+0.96*3+1,1.05) {$01$};
    	\node[circle, minimum height=1.1cm, dotted] (we) at (0.68+0.96*4+1,1.05) {$10$};
    	
    	\node[circle, minimum height=1.1cm, dotted] (we) at (-1.5,-0.1) {tAnn};
    	\draw[-, sloped] (-1.5+0.9,-1.5) -- node {}(-1.5+0.9, 1.3);
    	\draw[-, sloped] (-1.5+0.9,-1.5) -- node {}(-1.5+1.1,-1.5);
    	\draw[-, sloped] (-1.5+0.9,1.3) -- node {}(-1.5+1.1, 1.3);
    	
    	\draw[-, sloped] (-0.15+1,0.5) -- node {}(0.68+1,0.8);
    	\draw[-, sloped] (1.25+1,0.5) -- node {}(0.68+0.96*1+1,0.8);
    	\draw[-, sloped] (2.6+1,0.5) -- node {}(0.68+0.96*2+1,0.8);
    	\draw[-, sloped] (4.05+1,0.5) -- node {}(0.68+0.96*3+1,0.8);
    	\draw[-, sloped] (5.45+1,0.5) -- node {}(0.68+0.96*4+1,0.8);
    	
    	\draw[dotted] (0.2,1.6) rectangle (0.7,2.1) node[label={[label distance=0.25cm, minimum height=0cm]}, pos=.5] {$\bot$};
    	\draw[] (1.1,1.6) rectangle (1.6,2.1) node[label={[label distance=0.25cm, minimum height=0cm]}, pos=.5] {$c$};
    	\draw[] (2,1.6) rectangle (2.5,2.1) node[label={[label distance=0.25cm, minimum height=0cm]}, pos=.5] {$e$};
    	\node[circle, minimum height=1.1cm, dotted] (we) at (-1.5,1.85) {Stack};
    	
    	\draw[->, sloped] (1.1,1.85) -- node {}(0.7,1.85);
    	\draw[->, sloped] (2,1.85) -- node {}(1.6,1.85);
    
    	\draw[] (0.2,2.4) rectangle (0.7,2.9) node[label={[label distance=1.05cm, minimum height=0cm]}, pos=.5] {};
    	\draw[fill=black!30] (0.7,2.4) rectangle (1.2,2.9) node[label={[label distance=1.05cm, minimum height=0cm]}, pos=.5] {};
    	\node[circle, scale=0.2, draw, fill] (t1) at (0.45,2.65) {};
    	\node[circle, scale=0.2, draw, fill] (t2) at (0.95,2.65) {};
    	\node[circle, minimum height=1.1cm, dotted] (we) at (-1.5,2.65) {$top$};
    	
        \path[every node/.style={font=\sffamily\small}]
    	    (t1) edge[->] node [left] {} (0.45, 2.1)
    	    (t2) edge[bend left, ->] node [left] {} (2.25, 2.1);
    	    
        \node[circle, minimum height=1.1cm, dotted] (we) at (-1.5,3.45) {$cEpoch$};
        \draw[] (0.2,3.2) rectangle (0.7,3.7) node[label={[label distance=1.05cm, minimum height=0cm]}, pos=.5] {$10$};
        
    	\end{scope}
    	\end{tikzpicture}
		\subcaption{First configuration}
		\label{fig:steady}
	\end{subfigure}	
	\hspace{0.202cm}
	\begin{subfigure}[b]{0.7\textwidth}
    	\begin{tikzpicture}
    	\begin{scope}[scale=0.85, transform shape]
    	
    	\draw[] (-0.8+1,-1.5) rectangle (0.5+1,-0.5) node[label={[label distance=0.25cm, minimum height=0cm]}, pos=.5] {};
    	\draw[fill=black!30] (-0.8+1,-0.5) rectangle (0.5+1,0.5) node[label={[label distance=0.25cm, minimum height=0cm]}, pos=.5] {};
    	\draw[] (0.6+1,-1.5) rectangle (1.9+1,-0.5) node[label={[label distance=0.25cm, minimum height=0cm]}, pos=.5] {};
    	\draw[fill=black!30] (0.6+1,-0.5) rectangle (1.9+1,0.5) node[label={[label distance=0.25cm, minimum height=0cm]}, pos=.5] {};
    	\draw[] (2+1,-1.5) rectangle (3.3+1,-0.5) node[label={[label distance=0.25cm, minimum height=0cm]}, pos=.5] {};
    	\draw[fill=black!10] (2+1,-0.5) rectangle (3.3+1,0.5) node[label={[label distance=0.25cm, minimum height=0cm]}, pos=.5] {};
    	\draw[fill=black!30] (3.4+1,-1.5) rectangle (4.7+1,-0.5) node[label={[label distance=0.25cm, minimum height=0cm]}, pos=.5] {};
    	\draw[] (3.4+1,-0.5) rectangle (4.7+1,0.5) node[label={[label distance=0.25cm, minimum height=0cm]}, pos=.5] {};
    	\draw[fill=black!30] (4.8+1,-1.5) rectangle (6.1+1,-0.5) node[label={[label distance=0.25cm, minimum height=0cm]}, pos=.5] {};
    	\draw[] (4.8+1,-0.5) rectangle (6.1+1,0.5) node[label={[label distance=0.25cm, minimum height=0cm]}, pos=.5] {};
    	
    	
    	\node[circle, minimum height=1.1cm, dotted] (we) at (-0.15+1,0.3) {\footnotesize{$\func{Push}(g)$}};
    	\node[circle, minimum height=1.1cm, dotted] (we) at (1.25+1,0.3) {\footnotesize{$\func{Pop}()$}};
    	\node[circle, minimum height=1.1cm, dotted] (we) at (2.65+1,0.3) {\footnotesize{$\func{Push}(c)$}};
    	\node[circle, minimum height=1.1cm, dotted] (we) at (4.05+1,0.3) {\footnotesize{$\func{Pop}()$}};
    	\node[circle, minimum height=1.1cm, dotted] (we) at (5.45+1,0.3) {\footnotesize{$\func{Push}(e)$}};
    	
    	\node[circle, minimum height=1.1cm, dotted] (we) at (-0.15+1,0) {\footnotesize{$10$}};
    	\node[circle, minimum height=1.1cm, dotted] (we) at (1.25+1,0) {\footnotesize{$10$}};
    	\node[circle, minimum height=1.1cm, dotted] (we) at (2.65+1,0) {\footnotesize{$4$}};
    	\node[circle, minimum height=1.1cm, dotted] (we) at (4.05+1,0) {\footnotesize{$4$}};
    	\node[circle, minimum height=1.1cm, dotted] (we) at (5.45+1,0) {\footnotesize{$8$}};
    	
    	\node[circle, minimum height=1.1cm, dotted] (we) at (-0.15+1,-0.3) {\footnotesize{\ack}};
    	\node[circle, minimum height=1.1cm, dotted] (we) at (1.25+1,-0.3) {\footnotesize{$g$}};
    	\node[circle, minimum height=1.1cm, dotted] (we) at (2.65+1,-0.3) {\footnotesize{\ack}};
    	\node[circle, minimum height=1.1cm, dotted] (we) at (4.05+1,-0.3) {\footnotesize{$b$}};
    	\node[circle, minimum height=1.1cm, dotted] (we) at (5.45+1,-0.3) {\footnotesize{\ack}};
    	
    	\draw[-, anchor=north] (-0.8+1,-0.5) -- node[above left, scale=0.8, sloped] {} (0.5+1,-0.5);
    	\draw[-, anchor=north] (0.6+1,-0.5) -- node[above left, scale=0.8, sloped] {} (1.9+1,-0.5);
    	\draw[-, anchor=north] (2+1,-0.5) -- node[above left, scale=0.8, sloped] {} (3.3+1,-0.5);
    	\draw[-, anchor=north] (3.4+1,-0.5) -- node[above left, scale=0.8, sloped] {} (4.7+1,-0.5);
    	\draw[-, anchor=north] (4.8+1,-0.5) -- node[above left, scale=0.8, sloped] {} (6.1+1,-0.5);
    	
    	\node[circle, minimum height=1.1cm, dotted] (we) at (-0.15+1,-0.7) {\footnotesize{$\func{Push}(d)$}};
    	\node[circle, minimum height=1.1cm, dotted] (we) at (1.25+1,-0.7) {\footnotesize{$\func{Pop}()$}};
    	\node[circle, minimum height=1.1cm, dotted] (we) at (2.65+1,-0.7) {\footnotesize{$\func{Push}(a)$}};
    	\node[circle, minimum height=1.1cm, dotted] (we) at (4.05+1,-0.7) {\footnotesize{$\func{Push}(f)$}};
    	\node[circle, minimum height=1.1cm, dotted] (we) at (5.45+1,-0.7) {\footnotesize{$\func{Push}(h)$}};
    	
    	\node[circle, minimum height=1.1cm, dotted] (we) at (-0.15+1,-1) {\footnotesize{$6$}};
    	\node[circle, minimum height=1.1cm, dotted] (we) at (1.25+1,-1) {\footnotesize{$8$}};
    	\node[circle, minimum height=1.1cm, dotted] (we) at (2.65+1,-1) {\footnotesize{$2$}};
    	\node[circle, minimum height=1.1cm, dotted] (we) at (4.05+1,-1) {\footnotesize{$10$}};
    	\node[circle, minimum height=1.1cm, dotted] (we) at (5.45+1,-1) {\footnotesize{$10$}};
    	
    	\node[circle, minimum height=1.1cm, dotted] (we) at (-0.15+1,-1.3) {\footnotesize{\ack}};
    	\node[circle, minimum height=1.1cm, dotted] (we) at (1.25+1,-1.3) {\footnotesize{$d$}};
    	\node[circle, minimum height=1.1cm, dotted] (we) at (2.65+1,-1.3) {\footnotesize{\ack}};
    	\node[circle, minimum height=1.1cm, dotted] (we) at (4.05+1,-1.3) {\footnotesize{\ack}};
    	\node[circle, minimum height=1.1cm, dotted] (we) at (5.45+1,-1.3) {\footnotesize{$\bot$}};
    	
    	\node[circle, minimum height=1.1cm, dotted] (we) at (-0.15+1,-1.75) {$p_1$};
    	\node[circle, minimum height=1.1cm, dotted] (we) at (1.25+1,-1.75) {$p_2$};
    	\node[circle, minimum height=1.1cm, dotted] (we) at (2.65+1,-1.75) {$p_3$};
    	\node[circle, minimum height=1.1cm, dotted] (we) at (4.05+1,-1.75) {$p_4$};
    	\node[circle, minimum height=1.1cm, dotted] (we) at (5.45+1,-1.75) {$p_5$};
    	
    	\node[circle, minimum height=1.1cm, dotted] (we) at (-1.5+1.3,0.3) {\footnotesize{\func{Op}}};
    	\node[circle, minimum height=1.1cm, dotted] (we) at (-1.5+1.3,0) {\footnotesize{epoch}};
    	\node[circle, minimum height=1.1cm, dotted] (we) at (-1.5+1.3,-0.3) {\footnotesize{val}};
    	\node[circle, minimum height=1.1cm, dotted] (we) at (-1.5+1.3,-0.7) {\footnotesize{\func{Op}}};
    	\node[circle, minimum height=1.1cm, dotted] (we) at (-1.5+1.3,-1) {\footnotesize{epoch}};
    	\node[circle, minimum height=1.1cm, dotted] (we) at (-1.5+1.3,-1.3) {\footnotesize{val}};

    	\draw[] (0.2+1,0.8) rectangle (5+1,1.3) node[label={[label distance=2.65cm]}, pos=.5] {};
    	\draw[-, anchor=north] (0.2+0.96*1+1,0.8) -- node[above left, scale=0.8, sloped] {} (0.2+0.96*1+1,1.3);
    	\draw[-, anchor=north] (0.2+0.96*2+1,0.8) -- node[above left, scale=0.8, sloped] {} (0.2+0.96*2+1,1.3);
    	\draw[-, anchor=north] (0.2+0.96*3+1,0.8) -- node[above left, scale=0.8, sloped] {} (0.2+0.96*3+1,1.3);
    	\draw[-, anchor=north] (0.2+0.96*4+1,0.8) -- node[above left, scale=0.8, sloped] {} (0.2+0.96*4+1,1.3);
    	
    	\node[circle, minimum height=1.1cm, dotted] (we) at (-1.5+1.3,1.05) {\small{valid}};
    	
    	\node[circle, minimum height=1.1cm, dotted] (we) at (0.68+1,1.05) {$10$};
    	\node[circle, minimum height=1.1cm, dotted] (we) at (0.68+0.96*1+1,1.05) {$10$};
    	\node[circle, minimum height=1.1cm, dotted] (we) at (0.68+0.96*2+1,1.05) {$10$};
    	\node[circle, minimum height=1.1cm, dotted] (we) at (0.68+0.96*3+1,1.05) {$11$};
    	\node[circle, minimum height=1.1cm, dotted] (we) at (0.68+0.96*4+1,1.05) {$11$};
    	
    	\draw[-, sloped] (-0.15+1,0.5) -- node {}(0.68+1,0.8);
    	\draw[-, sloped] (1.25+1,0.5) -- node {}(0.68+0.96*1+1,0.8);
    	\draw[-, sloped] (2.6+1,0.5) -- node {}(0.68+0.96*2+1,0.8);
    	\draw[-, sloped] (4.05+1,0.5) -- node {}(0.68+0.96*3+1,0.8);
    	\draw[-, sloped] (5.45+1,0.5) -- node {}(0.68+0.96*4+1,0.8);
    	
    	\node[circle, minimum height=1.1cm, dotted] (we) at (-1.5,-0.1) {tAnn};
    	\draw[-, sloped] (-1.5+0.9,-1.5) -- node {}(-1.5+0.9, 1.3);
    	\draw[-, sloped] (-1.5+0.9,-1.5) -- node {}(-1.5+1.1,-1.5);
    	\draw[-, sloped] (-1.5+0.9,1.3) -- node {}(-1.5+1.1, 1.3);
    	
    	\draw[dotted] (0.2,1.6) rectangle (0.7,2.1) node[label={[label distance=0.25cm, minimum height=0cm]}, pos=.5] {$\bot$};
    	\draw[] (1.1,1.6) rectangle (1.6,2.1) node[label={[label distance=0.25cm, minimum height=0cm]}, pos=.5] {$c$};
    	\draw[] (2,1.6) rectangle (2.5,2.1) node[label={[label distance=0.25cm, minimum height=0cm]}, pos=.5] {$e$};
    	\draw[] (2.9,1.6) rectangle (3.4,2.1) node[label={[label distance=0.25cm, minimum height=0cm]}, pos=.5] {$f$};
    	\node[circle, minimum height=1.1cm, dotted] (we) at (-1.5,1.85) {Stack};
    	
    	\draw[->, sloped] (1.1,1.85) -- node {}(0.7,1.85);
    	\draw[->, sloped] (2,1.85) -- node {}(1.6,1.85);
    	\draw[->, sloped] (2.9,1.85) -- node {}(2.5,1.85);
    
    	\draw[] (0.2,2.4) rectangle (0.7,2.9) node[label={[label distance=1.05cm, minimum height=0cm]}, pos=.5] {};
    	\draw[fill=black!30] (0.7,2.4) rectangle (1.2,2.9) node[label={[label distance=1.05cm, minimum height=0cm]}, pos=.5] {};
    	\node[circle, scale=0.2, draw, fill] (t1) at (0.45,2.65) {};
    	\node[circle, scale=0.2, draw, fill] (t2) at (0.95,2.65) {};
    	\node[circle, minimum height=1.1cm, dotted] (we) at (-1.5,2.65) {$top$};
    	
        \path[every node/.style={font=\sffamily\small}]
    	    (t1) edge[->] node [left] {} (0.45, 2.1)
    	    (t2) edge[bend left, ->] node [left] {} (2.25, 2.1);
    	    
        \node[circle, minimum height=1.1cm, dotted] (we) at (-1.5,3.45) {$cEpoch$};
        \draw[] (0.2,3.2) rectangle (0.7,3.7) node[label={[label distance=1.05cm, minimum height=0cm]}, pos=.5] {$10$};
        
    	\end{scope}
    	\end{tikzpicture}

		\subcaption{Second configuration}
		\label{fig:combine}
	\end{subfigure}

	\caption{DFC: data structure illustrative example.}
	\label{fig:DFC-example}
\end{figure*}

%% file: Memory.tex
\section{Memory Management} \label{implementation}
\label{sec-memory}
Several general \emph{Persistent Memory Management} (\emph{PMM}) schemes have been proposed in recent years. Makalu \cite{Makalu} is a PMM for NVM-based systems. A PMM was also designed and implemented as part of Atlas \cite{Atlas}. 
However, for efficiency reasons, many applications designed for NVM-based systems, and specifically PTMs, implement their own memory management schemes. For example, Romulus \cite{correia2018romulus} and OneFile \cite{ramalhete2019onefile} pre-allocate all required memory before performing any transactions and maintain it during the execution with no extra calls to an external PMM. This allows the application to optimize memory management to fit its specific requirements, while avoiding any memory leakage in case of a crash.

Our implementation of DFC takes a similar approach and first allocates all required memory into a nodes pool of a user-defined size. During the execution of the algorithm, it utilizes this memory by synthetically allocating and deallocating the nodes of the actual data-structure (lines \ref{allocate-new-node}, \ref{deallocate}). Before exiting, this memory is freed. 
Therefore it is insignificant to optimize the single allocation and deallocation from the NVM at the beginning and the end of the execution, respectively. We used libpmemobj-cpp for doing that.


DFC uses several fixed-size non-volatile variables: $cEpoch$, $top$ and  the announcement array $tAnn$, whose number of entries equals the number of threads that may access the data structure.
The actual data-structure consists of nodes that represent the values that reside in the stack and which can be accessed via the $top$ pointers.
Node allocation is implemented by finding some \textit{free} node from the pool that was allocated beforehand, followed by a declaration that the node is \textit{used} using a flag bit.
Similarly, deallocating a node is implemented by resetting the \textit{used} flag of that node. 

To track which nodes are free and which are used, we implemented a standard bit-map hierarchy. Assuming a 64-bit system, we use 64 words, in which each bit indicates whether the corresponding node is currently used or not, for a total of $64^2$ nodes. An additional word is used at the ``root'' of this shallow tree, indicating which of the lower-level words have at least a single free node. Each time a node is allocated or deallocated, the root word and the lower-level word corresponding to the node are accessed. This hierarchy can be easily extended to support more nodes by adding more levels, at the cost of a moderate increase in the time complexity of memory management routines.

In order to support persistence in the presence of system crashes, we use the following approach. During the \func{Recover} procedure, the recovery combiner executes a \emph{garbage collection} (GC) cycle in line \ref{gc-recovery}, before accessing the memory. Notice that this GC cycle is performed once by the combiner, in the absence of concurrency, while all other threads are waiting for the combiner to release the lock. During the GC cycle, the combiner marks all nodes that are accessible from the active $top$ entry as \textit{used}, and all other nodes as \textit{free}. As a result, there is no need to keep the bit-map tree persistent and so we store it in volatile memory. Clearly, the nodes themselves must be kept persistent. This GC mechanism achives very lightweight memory management in normal operation at the cost of more expensive recovery. Since crashes are typically infrequent, this is a reasonable tradeoff.



%% file: Evaluation.tex
\section{Experimental Evaluation}
\label{sec-evaluation}

\newcommand{\DFC}{\mbox{\sc Dfc}}
\newcommand{\DFCTotal}{\mbox{\sc Dfc-Total}}
\newcommand{\Romulus}{\mbox{\sc Romulus}}
\newcommand{\OneFile}{\mbox{\sc OneFile}}
\newcommand{\PMDK}{\mbox{\sc PMDK}}

In order to evaluate the performance of the DFC stack, we compared it to implementations using Romulus \cite{correia2018romulus}, OneFile \cite{ramalhete2019onefile} and libpmemobj-cpp (PMDK) \cite{libpmemobjcpp-git}.
Romulus is a lock-based PTM that uses flat combining for update transactions. It uses two copies of the memory, a main copy and a backup copy. Transactions are first performed and persisted to the main copy, and then are copied to the backup. In case of a crash, one of the copies is consistent, and is the one used upon recovery. OneFile is a wait-free PTM. It uses  double-word-compare-and-swap (DCAS) instructions. Each update transaction is associated with a unique transaction identifier and all writes are done using a DCAS, writing the new value together with the identifier of the transaction that issues the write. PMDK is an open-source collection of libraries and tools to simplify managing and accessing persistent memory devices. It provides an undo-log based PTM. We note that none of these PTMs provides detectability.

\paragraph{Experimental Setting and Benchmarks.}
Our experiments were conducted using a machine 
equipped with two sockets of Intel Xeon Gold 5215 processors, 
each with a cache size of 13.75 MB. It contains a total of 20 physical and 40 logical cores and 4 256GB Intel Optane Persistent Memory Modules configured to the App Direct mode. The machine contains also 192 GB of RAM and runs on CentOS Linux with kernel version 5.8.7. All algorithms were implemented in C++ and compiled with g++ (version 9.1.0) with O2 optimizations.
On Intel architectures~\cite{Performance-Measurements}, the \pwb\ instruction is translated to \texttt{clwb}, \texttt{clflush} or \texttt{clflushopt} depending on what the machine at hand supports, where \texttt{clflushopt} is the most efficient option. In addition, the \pfence\ and \psync\ instructions are translated to \texttt{sfence}. Consequently, our code uses the \texttt{clflushopt} instruction and assumes that the execution of \texttt{sfence} ensures also the functionality of \psync.

The first benchmark we used is the \textit{push-pop} benchmark, in which we executed 1M couples of push and pop operations, that were distributed equally between the threads. In the second benchmark, called \textit{rand-op}, each thread chooses randomly and independently each of its operations to be either a push or a pop operation. In total, 2M operations are performed, distributed equally between the threads.
We ran the benchmarks using a varying number of threads in order to test algorithms' throughput and scalability. Each experiment was repeated 10 times and we report on median results.

\begin{figure*}[tb]
	\begin{subfigure}[c]{0.5\textwidth}
		\includegraphics[height=3.9cm]{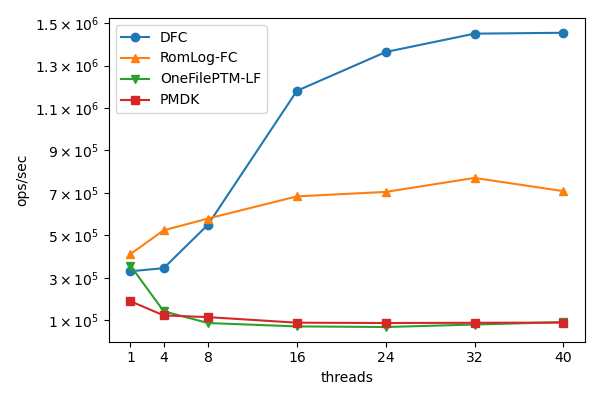}
		\subcaption{push-pop throughput}
		\label{fig:push-pop-throughput}
	\end{subfigure}		
	\begin{subfigure}[c]{0.5\textwidth}
		\includegraphics[height=3.9cm]{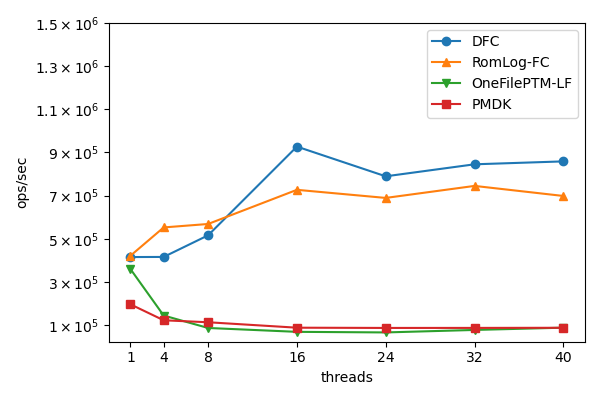}
		\subcaption{rand-op throughput}
		\label{fig:rand-op-throughput}
	\end{subfigure}		
	\begin{subfigure}[c]{0.5\textwidth}
		\includegraphics[height=4.1cm]{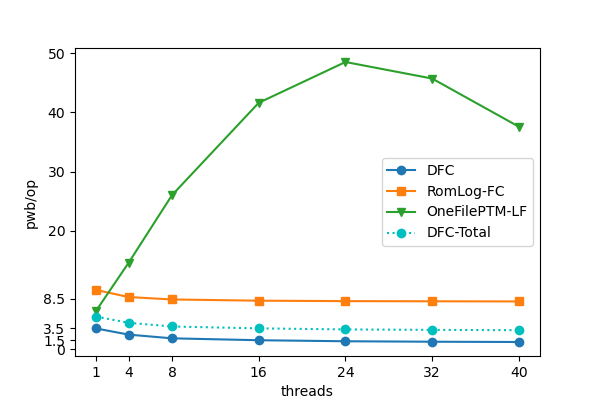}
		\subcaption{push-pop \#\pwb s}
		\label{fig:push-pop-pwb}
	\end{subfigure}		
	\begin{subfigure}[c]{0.5\textwidth}
		\includegraphics[height=4.07cm]{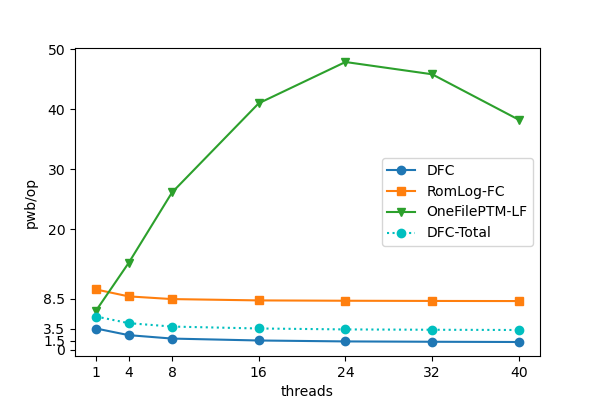}
		\subcaption{rand-op \#\pwb s}
		\label{fig:rand-op-pwb}
	\end{subfigure}		
	\begin{subfigure}[c]{0.5\textwidth}	
		\includegraphics[height=4.1cm]{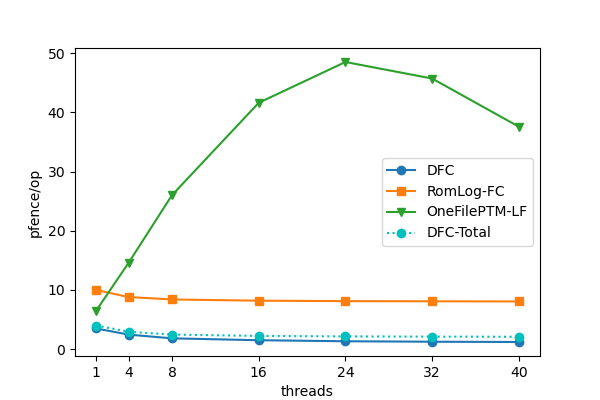}
		\subcaption{push-pop \#\pfence s}
		\label{fig:push-pop-pfence}
	\end{subfigure}
	\begin{subfigure}[c]{0.5\textwidth}	
		\includegraphics[height=4.07cm]{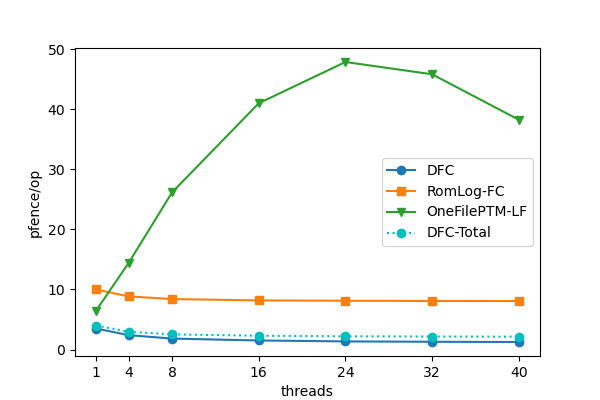}
		\subcaption{rand-op \#\pfence s}
		\label{fig:rand-op-pfence}
	\end{subfigure}	
	
	\caption{Throughput, \pwb s, \pfence s for push-pop and rand-op stack benchmarks.}
	\label{fig:graphs}
\end{figure*}

\paragraph{Analysis of Stack Experimental Results.}
Figure~\ref{fig:push-pop-throughput} shows that \DFC\ and \Romulus\ outperform \OneFile\ and \PMDK\ by a wide margin for all concurrency levels except 1. \DFC\ is outperformed by \Romulus\ for up to 8 threads. However, \DFC\ scales linearly from 4 to 16 threads and continues to scale up until concurrency levels of more than 30 threads. Specifically, with 16 threads, \DFC\ outperforms \Romulus, \OneFile\ and \PMDK\ by factors of $\times 1.727$, $\times 16.542$, $\times 13.289$, respectively. It maintains and even slightly increases these factors for 40 threads, to $\times 2.053$, $\times 15.868$ and $\times 16.279$, respectively.

Figures \ref{fig:push-pop-pwb}-\ref{fig:push-pop-pfence} present the average number of persistence instructions (\pwb\ and \pfence\ instructions, respectively) per operation performed by the algorithms across the concurrency range. It can be seen that algorithm performance is highly dependant on the number of persistence instructions executed by its operations.
As the \PMDK\ stack is significantly outperformed by all other algorithms in all concurrency levels, we do not analyze its performance in the evaluation that follows.\footnote{
See~\cite{correia2020persistent} for a similar \pwb\ evaluation of queue implementations,
including PMDK.}
\OneFile\ assumes an x86 architecture in which a compare-and-swap (CAS) instruction acts as an implicit \pfence\ and therefore does not execute \pfence\ instructions explicitly. Consequently, we count the number of CAS instructions in \OneFile\ as an estimate of the number of \pfence\ instructions it executes.
For the DFC stack, we present two types of measurements: \DFC\ in blue, and \DFCTotal\ in dashed blue. \DFCTotal\ refers to the total number of \pwb\ and \pfence\ instructions performed during the execution, while \DFC\ excludes the 
persistence instructions that are performed during the announcement process (lines \ref{parallel1}, \ref{parallel2}), since those are performed by threads in parallel.

The parallel \pwb\ and \pfence\ instructions performed by any specific thread do not block the progress of other threads, and therefore penalize the performance of the DFC stack less than persistence instructions performed by a combiner thread. This statement is supported by the fact that the throughput of \DFC\ outperforms that of \Romulus, although the \pfence\ count of \DFCTotal\ is higher than that of \Romulus, while the \pfence\ count of \DFC\ is lower.
Furthermore, we emphasize that although in general all \pwb\ instructions incur a similar cost, the execution time of a \pfence\ instruction may vary significantly. In fact, since \pfence\ instructions enforce the ordering and completion of all proceeding \pwb\ instructions, the execution time of each \pfence\ instruction highly depends on the number of \pwb\ instructions that precede it.
Thus, the costs of the \pfence\ instructions of \DFC\ (inside the \func{Combine} procedure) and \Romulus\ are more dominant than the costs of the additional two parallel \pfence\ instructions that are only preceded by a small number of \pwb\ instructions.

The \pwb\ and \pfence\ graphs have a lot in common. For both, in the sequential case, \OneFile\ starts with a small number of persistence instructions per operation compared to the other algorithms. However, for concurrency levels of 4 or more, it requires a larger number of persistency instructions. This explains the lack of  scalability in the throughput of \OneFile\ that is evident from Figure \ref{fig:push-pop-throughput}.
The persistency instruction counts of both \DFC\ and \Romulus\ start with a relatively steep descent as concurrency rises up to 16 threads and then stabilize. This explains the initial high throughput scalability ($\leq$ 16 threads) which is then followed by a more moderate throughput increase (for $>16$ threads) for both \DFC\ and \Romulus.
This similarity is not surprising, considering the fact that both \DFC\ and \Romulus\ employ flat combining, and consequently, for high levels of concurrency they are both able to combine together larger numbers of operation and thus reduce the average per-operation number of persistency instructions.
In addition, the ratios between the average number of \pwb s per operation between \Romulus\ and \DFCTotal\ and between \OneFile\ and \DFCTotal\ in concurrency level 40 ($\times 2.51$ and $\times 11.69$, respectively) are quite close to those between the corresponding throughputs. Indeed, as we have described, the number of per-operation \pwb\ instructions is an important indicator of an algorithm's performance in NVM-based platforms.

The push-pop benchmark is favorable to the DFC stack because it allows to combine many operations by eliminating pairs of \func{push} and \func{pop} operations. In order to evaluate DFC on a more challenging workload, we use the rand-op benchmark. 
All algorithms except \DFC\ present no significant change in throughput in comparison to the push-pop benchmark. However, in this benchmark, \DFC's performance drops significantly in comparison to the push-pop benchmark. Nevertheless, \DFC\ stack outperfors all other algorithms also in this benchmark for concurrency levels higher than 8, albeit by a significantly lower margin.
Figures~\ref{fig:push-pop-pfence} and~\ref{fig:rand-op-pfence} show that all algorithms maintain more-or-less 
the same average number of persistency instructions per operation on both benchmarks.

In order to better understand the performance drop in \DFC 's throughput in the rand-op benchmark, we also measured the average number of combining phases per operation in both. The results shown, a larger number of combining phases is performed by DFC in the rand-op benchmark. Consequently, a smaller number of operations is combined in each such phase on average, resulting in a smaller number of \func{push} and \func{pop} pairs that can be eliminated.




We remind the reader that \DFC\ is the only \emph{detectable}
stack implementation out of those we evaluated. In addition to providing higher throughput, the memory requirements of the DFC stack are roughly half of what is required by \Romulus . This is because whereas \Romulus\ uses two copies of all variables (including the main data structure, memory management data and required metadata), \DFC\ needs only a single copy of the stack and requires two copies of only the announcement structures and top pointers. 

\begin{figure*}[tb]
	\begin{subfigure}[c]{0.5\textwidth}
		\includegraphics[height=3.9cm]{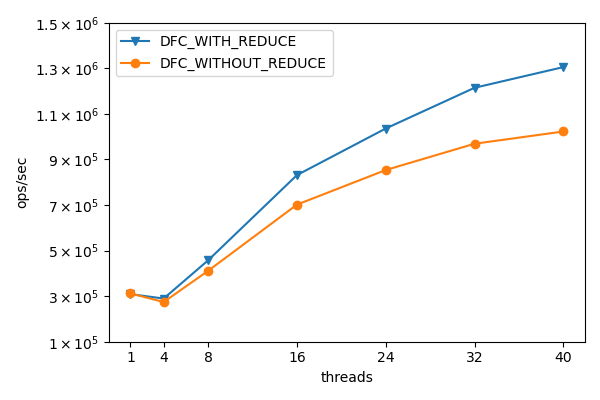}
		\label{fig:enq-deq-cancling}
	\end{subfigure}		
	\begin{subfigure}[c]{0.5\textwidth}
		\includegraphics[height=4.1cm]{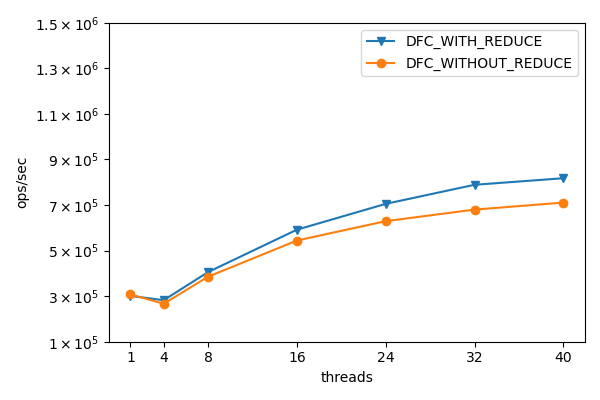}
		\label{fig:rand-op-canceling}
	\end{subfigure}

	\caption{Stack pair operations canceling, for push-pop (left) and rand-op (right).}
	\label{fig:cancle graphs}
\end{figure*}

Figure~\ref{fig:cancle graphs} shows the effect of canceling pair operations in the stack algorithm. 
In both benchmarks, canceling improved the stack performance,
however, the improvement is less significant in the rand-op benchmark,
due to less cancelling.

\begin{figure}[tb]
\centering
\resizebox{0.7\textwidth}{!}{
\includegraphics[]{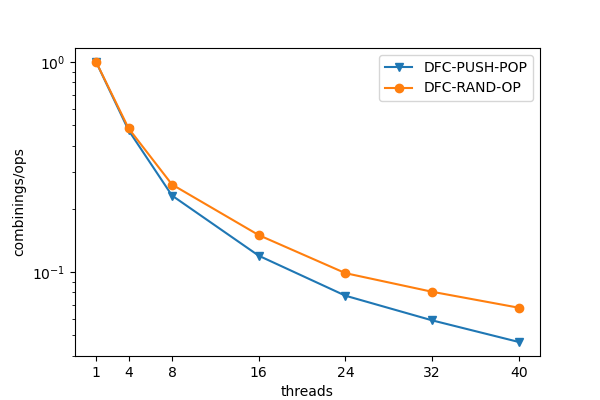}}
\caption{Number of DFC combining phases per operation. 
}
\label{fig:num-combine}
\end{figure}

The performance evaluation of our queue and deque implementations shows the same trends, 
and appears in Appendix~\ref{queue and deque}.

%% file: Discussion.tex
\section{Discussion}
\label{sec-discussion}

This work presents the DFC stack---the first non-transactional persistent stack algorithm. 
The new algorithm provides detectability and higher throughput in comparison to PTM-based alternatives, especially so in workloads where many \func{Push} and \func{Pop} operation-pairs can be eliminated. 
It establishes that the usage of FC can leverage object-specific semantics so that the number of persistence instructions is significantly reduced. 
We have also shown how DFC can applied to obtain additional detectable objects---queues and doubly-ended queues.

Friedman et al.~\cite{FriedmanHMP18} introduced detectability 
and proposed a detectable queue implementation. 
Detectability was formalized by \emph{nesting-safe recoverable 
linearizability}~\cite{attiya2018nesting}. An alternative formalization named \emph{detectable sequential specification}, recently proposed by Li and Golab \cite{DBLP:conf/podc/LiG21}, allows a thread to to declare, per object operation, whether it requires detectability or not, by extending the object's interface.
There exist detectable implementations for read-write registers 
and compare-and-swap (CAS)~\cite{attiya2018nesting, ben2019delay, Ben-BaruchHR20}. 
Capsules~\cite{ben2019delay} is a generic transformation making  
any implementation using read and CAS detectable, 
by replacing each primitive with its detectable counterpart. 
Persistent universal constructions~\cite{CohenGZ18, correia2020persistent} 
imply a persistent version of any concurrent object, 
but none of them is detectable. Very recent work \cite{DBLP:journals/corr/abs-2107-03492} presents detectable implementations of several objects, including a stack. Their algorithms employ combining as well, but, as they mention, there are some differences in technical implementation details. We have provided our previously-archived code to the authors and the performance evaluation they present shows higher throughput in comparison to our DFC stack for high concurrency levels, on a different machine than the one we used. We did not have access to their code, and therefore were not able to fully explore the differences between the two implementations nor to evaluate the performance of their implementation on our machine.